\documentclass[aps,prd,superscriptaddress,nofootinbib,twocolumn,floatfix]{revtex4}

\usepackage{slashed}
\usepackage{hyperref}
\usepackage{amsmath,amsfonts,amssymb}
\usepackage{booktabs}
\usepackage{comment}
\usepackage{siunitx}
\usepackage{xcolor}
\usepackage{verbatim}
\usepackage[english]{babel}

\usepackage{float}
\usepackage{tikz}

\usepackage{tikz-feynman}
\tikzfeynmanset{compat=1.1.0}

\usepackage{graphicx}

\newcommand{\be}{\begin{equation}}
\newcommand{\ee}{\end{equation}}
\newcommand{\bea}{\begin{equation}\begin{aligned}}
\newcommand{\eea}{\end{aligned}\end{equation}}

\newcommand{\gsim}{\lower.7ex\hbox{$\;\stackrel{\textstyle>}{\sim}\;$}}
\newcommand{\lsim}{\lower.7ex\hbox{$\;\stackrel{\textstyle<}{\sim}\;$}}

\begin{document}

\title{Beyond oblique contributions: $\Delta$r and W-mass in models with scalar leptoquarks}

\author{Carlo Marzo}
\affiliation{Laboratory for High Energy and Computational Physics, NICPB, R\"{a}vala 10, Tallinn 10143, Estonia}
\date{\today}

\begin{abstract}
We study the dominant non-universal contributions to muon decay for the five scalar leptoquarks (LQs) allowed by the Standard Model gauge symmetries.
By using a common computational framework, we isolate the new-Physics (NP) one-loop corrections over the observable $\Delta r$ and explore the connection with the $W$ boson mass. We highlight the role of non-universal terms considering, for the $SU(2)$ doublet and triplet LQs, a completely degenerate scenario, thus cancelling/moderating the effects of the oblique parameters S,T and U . We determine simple formulas that exhibit the leading non-universal behaviour generated by the introduction of extra Yukawa terms, and confront them against a present tension in $M_W$ and the projected FCC-ee accuracy.   
\end{abstract}

\maketitle

\vspace{0.3cm}

\section{Introduction}

The profound harmony between theoretical predictions and experimental probing clearly indicates how the Standard Model (SM) architecture can efficiently account for particle dynamics up to the Electroweak (EW) scale.  
A sheer consequence of this is that the primary research for direct production of new possible states, beyond the SM ones, must often rely on a growing scale of new experimental facilities in order to access much higher energies. A different path to unveil NP considers the narrowing of the experimental uncertainty so as to put on test the theoretical predictions beyond the main leading terms, the latter expected to be SM-dominated. A celebrated example is the $\rho$ parameter \cite{Veltman:1977kh}
\bea
\rho = \frac{M_W^2}{M_Z^2 \cos^2 \theta_W} \, ,
\eea
predicted to be 1 in the SM, a value backed by experiments to high accuracy. The small deviations from the $\rho_{SM} = 1$ benchmark are, hence, accounted for by radiative corrections, constraining the parameters entering at the loop level.
Similarly, deviations from \emph{natural} relations can be used to indirectly parametrize the radiative perturbation from them, when paralleled by precise theoretical determinations. 
It is a puzzling and exciting result that the standing most convincing signals of NP are indeed revealed by this accurate comparison. This is certainly the case for the measure of the anomalous magnetic moment of the muon 
\cite{Muong-2:2021ojo,Muong-2:2006rrc,Aoyama:2020ynm,Czarnecki:2002nt,Gnendiger:2013pva,Keshavarzi:2018mgv,Colangelo:2018mtw,Kurz:2014wya,Melnikov:2003xd,Athron:2021iuf}
that shows the large deviation 
\bea
\Delta a_{\mu} = (2.51 \pm 0.59) \times 10^{-9} \, ,
\eea
and points to a $4.2 \sigma$ difference from the SM. On top of this, the introduction of the W boson mass in the set of precisely determined observables, through the last probing by the CDF collaboration \cite{CDFII}
\bea
M_W^{\text{CDF II}} = 80433.5 \pm 94 \text{  MeV} \, , 
\eea
illuminated a new tension at the $7 \sigma$ level when compared with the SM predicted value of \cite{Haller:2018nnx} 
\bea
M_W^{2018} = 80354 \pm 7 \text{  MeV}\, .
\eea
In light of this and the appealing future prospect of a reduction to $0.5$ MeV for the experimental error \cite{FCC:2018byv}, we compute the impact of a particular class of extensions of the SM to the predicted value of $M_W$. Differently from most of the solutions proposed so far (an incomplete list being \cite{Strumia:2022qkt,Asadi:2022xiy,Primulando:2022vip,Bahl:2022gqg,CentellesChulia:2022vpz,Kawamura:2022fhm,Chowdhury:2022dps,Isaacson:2022rts,He:2022zjz,Dcruz:2022dao,Kim:2022xuo,Barman:2022qix,Wang:2022dte,Botella:2022rte,Gupta:2022lrt,Zhou:2022cql,Cai:2022cti,Batra:2022pej,Abouabid:2022lpg,Heeck:2022fvl,Addazi:2022fbj,Cheng:2022aau,Batra:2022org,Borah:2022zim,Cao:2022mif,Bhaskar:2022vgk,Baek:2022agi,Chowdhury:2022moc,Kanemura:2022ahw,Popov:2022ldh,Kawamura:2022uft,Ghoshal:2022vzo,Ahn:2022xax,Balkin:2022glu,Biekotter:2022abc,Paul:2022dds,Babu:2022pdn,Gu:2022htv,DiLuzio:2022xns,Sakurai:2022hwh,Athron:2022qpo,Athron:2022isz,Han:2022juu,Basiouris:2022wei}), which determine $M_W$ by considering the universal corrections to the muon decay, encoded in the oblique parameters of \cite{Peskin:1991sw}, we focus on the \emph{non-oblique} corrections, which depend on the non-universal, flavor-dependent modification of the SM. As a targeted extension of the SM we tackle the scalar leptoquark case \cite{Dorsner:2016wpm}, which is often invoked with peculiar flavour textures, as a candidate solution to resolve SM/experiment tensions \cite{Angelescu:2018tyl,Crivellin:2022mff,Crivellin:2019dwb,Crivellin:2019szf,Allanach:2022iod,Sahoo:2021vug,Belanger:2021smw,Nomura:2021oeu,Angelescu:2021lln,Altmannshofer:2020ywf,Aydemir:2019ynb,DiLuzio:2017vat,Cai:2017wry,Popov:2016fzr,Li:2016vvp,
Becirevic:2016yqi,Bauer:2015knc,DelleRose:2020qak,Dedes:2021abc,ColuccioLeskow:2016dox,Crivellin:2020tsz,Crivellin:2020ukd}.
\section{Scalar LQ models}
The colourless Higgs sector of the Standard Model (SM) forbids Yukawa-type interactions with a single field charged under the $SU(3)$ gauge group. Consequently, we have the emergence of global symmetries separately selecting the leptons/antileptons and quarks/antiquarks abundances in (perturbative) SM processes. The conservation of the leptonic ($\bold{L}$) and baryonic ($\bold{B}$) number is not enforced by any fundamental consistency requirement and is not expected to survive at higher energies, where interactions are selected by the less strict conservation of the \emph{fermion number} $\bf{F}$ $= 3$ $\bf{B} + \bf{L}$.

Within the restricted space of the SM gauge group, the possible ways to connect leptons and quarks are limited. In particular, when considering a single extension of the SM via scalar field, only five scenarios are possible. The direct interactions with a quark field only allow building invariant combinations out of the $\bf{3}$ or $\bar{\bf{3}}$ fundamental representations of $SU(3)$. Instead, both quarks and leptons experience, through their left-chirality components, the $SU(2)_{EW}$ gauge interaction and, therefore, more freedom is given in generating an invariant Yukawa sector with a coloured scalar. We follow throughout this work, and we refer to, in order to disentangle the index structure of the interactions, the conventions and definitions of \cite{Dorsner:2016wpm}.

\subsubsection*{$SU(2)_{EW}$ singlet}

Under the SM gauge group $SU(3) \times SU(2)_{EW} \times U(1)_Y$ two distinct leptoquark 
\emph{scalar} representations are possible, conventionally labeled $S_1$ and $\tilde{S}_1$ and transforming under $(\bf{\bar{3}},\bf{1},1/3)$ and, respectively, $(\bf{\bar{3}},\bf{1},4/3)$.
The allowed Yukawa terms for the $S_1$ case are consequently
\bea \label{LS1}
& \mathcal L_{S_1} =  \mathcal Y_{LL} \,\,  \bar Q_L^C \, S_1 \, \epsilon \,L_L + \mathcal Y_{RR} \,\,  \bar u_R^C \, S_1 \, e_R +  \\ 
& + \mathcal Z_{LL} \,\, \bar Q_L^C \, S_1^* Q_L + \mathcal Z_{RR} \,\, \bar u_R^C \, S_1^* d_R + \text{h.c.} \, ,
\eea
\\
and 
\bea \label{LS1t}
\mathcal L_{\tilde{S}_1} =  \mathcal Y_{RR} \,\,  \bar d_R^C \, \tilde{S}_1 \, e_R + \mathcal Z_{RR} \,\, \bar u_R^C \, \tilde{S}_1^* u_R + \text{h.c.} \,
\eea
for $\tilde{S}_1$. We kept a minimalistic notation by concealing the contractions of flavour, colour and $SU(2)_{EW}$ indices. These can be read off, when in doubt, from \cite{Dorsner:2016wpm}. It is intended that $\mathcal{Y}$ and $\mathcal{Z}$ are complex $3 \times 3$ matrices contracted with the flavour indices of the corresponding quark/lepton fields. In this work we do not consider a generation index for the LQs. In general, the diquark interactions $\mathcal Z$, which plays a critical role in curbing proton decay \cite{Dorsner:2016wpm}, can be neglected when computing, at one-loop, the low-energy observables under investigation.
It might appear that the rotation to mass eigenstates, triggered by spontaneous symmetry breaking (SSB) in the Higgs sector,
\bea
& u_L = \mathcal{O}^u_L \cdot u'_L\, , d_L = \mathcal{O}^d_L \cdot d'_L\, , u_R = \mathcal{O}^u_R \cdot u'_R, \\
& d_R = \mathcal{O}^d_R \cdot d'_R\, , e_L = \mathcal{O}^e_L e'_L, e_R = \mathcal{O}^e_R e'_R \, ,
\eea
would exacerbate the already convoluted flavour texture. Nevertheless, all rotations but one, in the absence of right-handed neutrino counterparts, can be conveyed in a field redefinition that leaves the kinetic term invariant. The precise ways to exploit such invariance are matters of taste. We adopt the choice that leads to a diagonal up-Yukawa in the Higgs sector which, in turn, reveals the quark mixing CKM matrix $V_{CKM}$ in the down sector via $Q_L  = \left( u_L, V_{CKM} d_L \right)$. 
We also drop the apostrophe in labelling mass eigenstates to lighten the notation. 
The CKM matrix can easily be kept in full during our computation and will affect the output only in the two cases $S_1$ and $S_3$. It is apparent to us that the gain in precision, obtained by keeping all terms in $V_{CKM}$, is outmatched by the gain in simplicity when we neglect its subdominant components. In light of this, we proceed by only including the main Cabibbo mixing
\bea
V_{CKM} = 
\begin{pmatrix}
\cos{\theta_{C}}  & \sin{\theta_{C}} & 0\\
- \sin{\theta_{C}}  & \cos{\theta_{C}} & 0\\
0  & 0 & 1 
\end{pmatrix} \, \, .
\eea
These scalar representations, in particular when considering also their diquark Yukawa terms, have a definite value of the global fermion number $\bf{F} = -2$, which is connected with the same-chirality type of interactions mediated by them.  
\subsubsection*{$SU(2)_{EW}$ doublet} \label{LQsDoublet}
There are two leptoquark representations that transform as doublets under $SU(2)_{EW}$,  
$R_2 \sim (\bf{3},\bf{2},7/6)$ and $\tilde R_2 \sim(\bf{3},\bf{2},1/6)$.
After SSB their components have a definite value of QED charge and can be defined as
\bea \label{LQsDoubletF}
R_2 = 
\begin{pmatrix}
R_2^{5/3} \\
R_2^{2/3} 
\end{pmatrix} \,\, , \,\,\,
\tilde{R}_2 = 
\begin{pmatrix}
\tilde R_2^{2/3} \\
\tilde R_2^{-1/3} \,  
\end{pmatrix}\,\, .
\eea
Their Yukawa interactions are 
\bea \label{LR2}
& \mathcal L_{R_2} =  -\mathcal Y_{RL} \,\,  \bar u_R \, R_2 \, \epsilon \,L_L + \mathcal Y_{LR} \,\,  \bar e_R \, R_2^* \, Q_L + \text{h.c.} \, ,
\eea
and 
\bea \label{LRt2}
& \mathcal L_{\tilde{R}_2} =  -\mathcal Y_{RL} \,\,  \bar d_R \, \tilde R_2 \, \epsilon \,L_L + \text{h.c.} \, .
\eea
The chirality-flipping interactions assign therefore $\bf{F} = 0$ to both doublet representations. 
\subsubsection*{$SU(2)_{EW}$ triplet}
Finally, we have the triplet LQ representation $S_3 \sim (\bf{3},\bf{3},1/3)$ containing the related triplet of charged scalars as in 
\bea
S_3 = 
\begin{pmatrix}
\left(S_3^{-4/3} + S_3^{2/3}\right)/\sqrt{2} \\
\left(S_3^{-4/3} - S_3^{2/3}\right)/ i \sqrt{2}\\
S_3^{-1/3}
\end{pmatrix} \,\, .
\eea
They are connected to the SM fermions sector via 
\bea \label{LS3}
& \mathcal L_{S_3} =  \mathcal Y_{LL} \,\,  \bar Q_L^C  \, \epsilon \, \left(\vec{\tau} \cdot \vec{S}_3\right)\,L_L +  \\
& \quad + \mathcal Z_{RR} \,\,  \bar Q_L^C \, \epsilon \, \left(\vec{\tau} \cdot \vec{S}_3\right)^{\dagger} \, Q_L  + \text{h.c.} \, ,
\eea
thus also defining a LQ of fermion number $\bf{F} = -2$.
\subsubsection*{LQ potential and SSB}
On top of gauge and Yukawa interactions, the LQs scalars affect also the Higgs potential which is modified by the extra terms
\bea
V_{LQ} = + M_{\mathcal{S}}^2 \bar{\mathcal{S}} \mathcal{S} + \lambda_{p} \bar{H} H \bar{\mathcal{S}} \mathcal{S} + \frac{\lambda_{LQ}}{4} \bar{\mathcal{S}} \mathcal{S} \bar{\mathcal{S}} \mathcal{S} \, . 
\eea
where $\mathcal{S}$ is one of the five representations discussed $S_1, \tilde{S}_1, R_2, \tilde{R}_2$ and $S_3$. The interplay between $\lambda_p$ and the explicit mass term $M_{\mathcal{S}}^2$ is responsible for the mass splitting between the different components of the doublet and the triplet scalar LQs. In our computation, we will focus exclusively on the Yukawa contributions and consider the extra quartics to be always negligible. The mass splitting can therefore, in first approximation, be neglected as well, so that $M_{\mathcal{S}}$ will represent the shared mass of the propagating LQs states.

\section{$\Delta$r, muon decay and the mass of the W}
\begin{figure}
    \centering
 \includegraphics[scale=0.3]{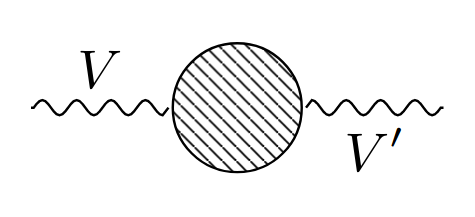}
\caption{Vector boson two-point functions involved in the OS renormalization of the SM. $V V'$ stands for $WW$, $ZZ$, $\gamma \gamma$ and $\gamma Z$. }
    \label{2VV}
\end{figure}

The determination of the $W$ mass in the SM is one of the earliest and most evident manifestations of its renormalizability and predictivity. To briefly review its peculiar role among the other EW parameters, we have to consider the main steps needed to confront the (UV) cut-off dependent predictions of the renormalizable SM, with the finite, physical observables produced by experiments.    
A standard way to proceed towards a smoother comparison is to perform a trade of the SM Lagrangian parameters 
\bea \label{unbr}
g_1, g_2, \mathcal Y_u, \mathcal Y_d, \mathcal Y_{e}, v_{\text{\tiny SM}}, \lambda, \mu^2
\eea
for the set 
\bea \label{bro}
M^2_Z, M^2_W, e_{\small{\text{\tiny QED}}}, m_{u_i}, m_{d_i}, m_{lep}, M^2_H, t\, .
\eea  
In \ref{unbr} a common labelling convention has been followed, with $g_1$ and $g_2$ being the Hypercharge and Weak coupling, $\mathcal Y_u, \mathcal Y_d$ and $ \mathcal Y_{e}$ the $3\times 3$ SM Yukawa matrices and the remaining parameters describing the Higgs potential.
Then, renormalizability grants that the UV dependence of the loop contributions can be completely concealed in opportune counterterms introduced by the redefinition
\bea \label{CoRen}
& M^2_Z \rightarrow M^2_Z  + \delta M^2_Z, \,\, M^2_W \rightarrow M^2_W  + \delta M^2_W,  \\
& m_f \rightarrow m_f  + \delta m_f, \quad (f = u_i,d_i, e,\mu, \tau)  \\
& M^2_H \rightarrow M^2_H  + \delta M^2_H, \,\, t \rightarrow t  + \delta t \, ,  \\
& e_{\text{\tiny QED}} \rightarrow e_{\text{\tiny QED}}\left(1 + \delta e \right) \,,
\eea
and field rescaling 
\bea \label{FiRen}
& W_{\mu} \rightarrow \left( 1 + \frac{\delta Z_W}{2}\right)W_{\mu}  ,  \\
& Z_{\mu} \rightarrow \left( 1 + \frac{\delta Z_Z}{2}\right)Z_{\mu} + \frac{1}{2}\delta Z_{ZA} A_{\mu},  \\ 
& A_{\mu} \rightarrow\left( 1 + \frac{\delta Z_A}{2}\right)A_{\mu} + \frac{1}{2}\delta Z_{AZ} Z_{\mu},  \\ 
& \Phi \rightarrow \left( 1 + \frac{\delta Z_H}{2}\right)\Phi , \quad (\Phi = H, \phi_0, \phi^{\pm}) \\
& \psi \rightarrow \left( 1 + \delta Z_{\psi}\right)\psi, \quad (\psi = l_L, l_R, \nu_L) \, ,
\eea
the latter ensuring also the finiteness of off-shell Green functions.
In the on-shell scheme (OS), in particular in the form adopted in \cite{Denner:1991kt}, renormalization is accomplished via a particular set of conditions: with the exception of the two parameters $e_{\text{\tiny QED}} $ and $t$, connected with the determination of the QED coupling strength and SM vacuum \cite{Denner:1991kt,Sirlin:1980nh}, the remaining set is identified with the pole masses of the corresponding two-point functions \ref{2VV}. 
The predictive power of the SM is maximised when its renormalized parameters are fixed by the most precise experimental inputs available. For the case of the $Z$ vector boson, the LEP exploration has indeed provided extremely accurate details of its production features, thus electing $M_Z$ as a reliable input. The same is true for the determination of $\alpha_{\text{\tiny QED}}$, whose counterterm in \ref{CoRen} can be directly connected to the precisely measured Thomson limit of the $e + \gamma \rightarrow  e$ scattering.     
The direct production of the $W$ boson at LEP and Tevatron has also yielded an experimental measure of its mass up to a precision of order $\sim 10$ MeV. Nevertheless, other experimental probes of the SM can compete and surpass this accuracy and represent a more favourable input. 
This was exploited in \cite{Sirlin:1980nh} within the so-called $G_F$ scheme, which elected as main EW inputs $M_Z$, $\alpha_{\text{\tiny QED}}$ and the Fermi constant $G_F$, linked to the muon decay. This choice frames the value of the $W$ mass as a predicted parameter 
\bea
M_W = M_W(M_Z, \alpha_{\text{\tiny QED}}, G_F, M_H, m_{top} \cdots)\, ,
\eea
indirectly determined through its role in the radiative corrections of the process $\mu \rightarrow e \nu_{\mu} \bar \nu_e$  

\subsection*{Radiative corrections to muon decay}
The value of the Fermi constant $G_F$ is fixed by the muon lifetime undressed from the QED contribution. Therefore, a connection to the measured value of $\tau_{\mu}$ is established in a theory where the renormalizable QED interaction is paired with the four-fermion contact term
\bea
\mathcal{L}_{Fermi} = -\frac{G_F}{\sqrt{2}} \left(\bar{\psi}_{\mu} \gamma^{\rho} P_L \psi_{\nu_{\mu}}\right) \left(\bar{\psi}_{e} \gamma_{\rho} P_L \psi_{\nu_{e}}\right)  +  h.c. \,\,\,.
\eea
In such an effective model the corresponding muon decay width is given by the relation
\bea \label{FermiMu}
\frac{1}{\tau_{\mu}} = \frac{G_F^2 m_{\mu}^2}{192 \pi^3}\left(1 + \delta_{QED}\right),
\eea
and the precision in the determination of $G_F$ increases with our ability of computing higher-order terms in $\delta_{QED}$. When switching to the renormalizable setup of the SM, the same quantity can be computed in the limit of zero external momentum and negligible lepton masses, giving
\bea \label{SMMu}
\frac{1}{\tau_{\mu}} = \frac{m_{\mu}^2}{192 \pi^3}\frac{g_w^4}{32 M_W^4}\left(1 + \delta^{SM}_{\mu}\right).
\eea
In \ref{SMMu}, $\delta^{SM}_{\mu}$ stands for the full set of radiative corrections to the muon decay originating from the SM interactions and particle content. When comparing the effective \ref{FermiMu} with the full theory forms \ref{SMMu}, we arrive at
\bea \label{GF}
&\frac{G_F}{\sqrt{2}} = \frac{g_w^2}{8 M_W^2} \left(1 + \frac{1}{2}\left( \delta^{SM}_{\mu} - \delta_{QED} \right)\right) 
 \\
&\quad = \frac{g_w^2}{8 M_W^2} \left(1 +\tilde{\delta}^{SM}_{\mu}\right)\, , 
\eea
which connects the degrees of freedom that shape $\tilde{\delta}^{SM}$ with $G_F$. The subtraction of the QED corrections from $\delta^{SM}_{\mu}$, in particular from EW box diagrams, is a non-trivial process and details of the procedure can be found in \cite{Green:1980bd,Bardin:1999ak}. For our purposes, it is sufficient to focus on the relation between the SM formula and the main deviation coming from new Physics. Eq.~\ref{GF} is usually rewritten in the form
\bea \label{MWmain}
& \left(1 - \frac{M^2_W}{M^2_Z} \right) \frac{M^2_W}{M^2_Z} = \frac{\pi \alpha}{\sqrt{2} G_F M_Z^2} \left( 1 + \Delta r \right)  \\
& \xrightarrow{RG} \frac{\pi \alpha}{\sqrt{2} G_F M_Z^2} \frac{1}{1 + \Delta r} ,  \hspace{1.2cm}  (\alpha = e^2/4\pi) 
\eea
with the last step accounting for the renormalization group resummation of large $\ln M_Z/m_{lep}$ terms \cite{Bardin:1999ak}.
Formula \ref{MWmain} is the equation that implicitly fixes the value of $M_W$ which appears both explicitly on the left-hand side as well as in the radiative corrections encoded in the parameter $\Delta r$. Within the SM, $\Delta r$ is completely known up to two \cite{Freitas:2002ve,Awramik:2003rn,Awramik:2003ee,Awramik:2002wv,Awramik:2002vu,Awramik:2002wn,Freitas:2002ja,Freitas:2000gg,Onishchenko:2002ve} and leading three \cite{Avdeev:1994db,vanderBij:2000cg,Chetyrkin:1996cf,Chetyrkin:1995js,Chetyrkin:1995ix,Boughezal:2004ef} and four loops \cite{Chetyrkin:2006bj,Boughezal:2006xk}. 
Such computations have allowed different predictions in the SM, for our purposes we mention two. 
Supported by the LHC measurements of the Higgs mass \cite{ATLAS:2012yve,CMS:2012qbp} the value $M_{W}^{2013} = 80.361$ GeV is derived in \cite{Heinemeyer:2013dia}. Different sources of errors can be attached to such estimate, the larger coming from the uncertainty in $m_t$ and accounting for shifts of $\pm 6$ MeV. Similarly, a close determination is also provided by electroweak fits \cite{Haller:2018nnx}, also adopting the same up-to-date theoretical formulas for the $W$ mass, and pointing to a value $M_{W}^{2018} = 80.354$ GeV and an error $\pm 7$ MeV. 
Deviations from the SM estimate are linked to deviations in $\Delta r = \Delta r^{SM} + \Delta r^{NP}$ and cause, with good approximation \cite{Lopez-Val:2014jva}, a shift in $M_W^{SM}$ of form
\bea \label{MWNP}
M_W^{NP} \sim M_W^{SM}\left( 1 -\frac{1}{2} \frac{s^2_W}{c^2_W-s^2_W}  \Delta r^{NP} \right) \, ,
\eea
where, at all orders in the OS scheme, $c_W = M_W/M_Z$ and $s^2_W = 1 - c_W^2 $. 
\subsection*{Computation of $\Delta r^{NP}$}
To assess the UV-finite value of $\Delta r^{NP}$ in the NP models under analysis, as well as for generality and future next-to-leading order applications, we have performed the OS renormalization for the SM parameters in \ref{bro}. This process has been assisted by an implementation of the models in \texttt{FeynRules} \cite{Christensen:2008py, Alloul:2013bka}, revisiting the lagrangians provided in \cite{Dorsner:2018ynv}, and an explicit determination of the counterterms in \ref{CoRen} \ref{FiRen} after imposing OS renormalization conditions \cite{Denner:1991kt}. Moreover, the computation has been assisted by the topological generator \texttt{FeynArts}, by \texttt{FormCalc} for the reduction to one-loop scalar integrals and rapid check of UV-finiteness \cite{Hahn:2000jm,Hahn:2000kx,Hahn:2016ebn}, and the analytical library \texttt{Package-X} \cite{Patel:2016fam} for the explicit form of the scalar loop integrals.
After rewriting the parameters of the unbroken phase \ref{unbr} in terms of \ref{bro}, and performing the shifts \ref{CoRen} \ref{FiRen}, we can write a generic formula for $\Delta r$ by explicitly computing the counterterms diagrams. We isolate the different terms by writing $\Delta r$ in the form
\bea \label{deltar}
&\Delta r = \Delta r^{\text{self-energies}}_{LQ} + \Delta r^{\text{triangles}}_{LQ} + \Delta r^{\text{boxes}}_{LQ} = \\
& = \tilde \Delta r^{\text{self-energies}}_{LQ} + \tilde \Delta r^{\text{triangles}}_{LQ} + \Delta r^{\text{boxes}}_{LQ}  +  \\ 
&+ 2 \delta e + \frac{\delta M_Z^2}{M_Z^2} - \frac{\delta M_W^2}{M_W^2} - \frac{\delta M_Z^2-\delta M_W^2}{M_Z^2-M_W^2} + \\ 
& + \frac{1}{2}\left(\delta Z_{e_{L}} + \delta Z_{{\mu}_{L}} + \delta Z_{\nu_{e, L}} + \delta Z_{\nu_{\mu, L}} \right) \, ,
\eea
where we use the notations $\Delta$ ($\tilde \Delta$) to label the renormalized (unrenormalized) type of diagrammatic contribution.  
This formula generalizes the parallel one derived in the context of oblique parametrization of EW processes \cite{Peskin:1991sw}, considering NP effects beyond those in vacuum polarization diagrams. Moreover, the validity of \ref{deltar} is not limited to the case of heavy NP.  
\subsection*{$\Delta r^{NP}$ and LQ models}

The chiral structure of the LQ interactions might introduce extra powers of the external leptonic masses causing a strong suppression of their contribution to the  muon decay. This is the case of the $\tilde S_1$ LQ, whose effect at one-loop is proportional to either $m_e$ or $m_{\mu}$ and can be completely ignored. For all the remaining cases we find that the Yukawa interactions might trigger non-negligible contributions to $\Delta r$, potentially competing in size with the universal counterparts. In the spirit of the oblique EW parameterization of NP effects over $\Delta r$, we also seek simplified formulas efficiently accounting for deviations from universality. This is not a granted result, given that even in the approximations standardly adopted in such computation, formulas coming from triangle and box diagrams can have an intricate structure.
Upon an expansion for large LQ masses, justified when $ M_{\mathcal S} \gg m_t,M_Z$, we find that box diagrams contribution simplifies to 
\bea \label{simpBOX}
& \Delta r^{\text{boxes}}_{LQ} = B \left(M_W,M_Z\right) \times \frac{\mathcal Y^4}{ M_{\mathcal S}^2} = \\ 
& =  \left(\frac{3}{ 32 } \frac{M_W^2}{M_Z^2} \frac{M_Z^2-M_W^2}{e^2 \, \pi^2}\right) \times \frac{\mathcal Y ^4}{ M_{\mathcal S}^2} \, .
\eea
In \ref{simpBOX} we have introduced the master \emph{box} function $B \left(M_W,M_Z\right)$ and used the shortcut $\mathcal Y^4$ to suggest a general quartic polynomial in the Yukawas. 
One-loop vertex corrections generate, instead, a more lengthy form for their leading contribution, which explicitly involves all the quark masses the LQ interact with. A simple numerical survey confirms the intuitive hypothesis that only third-generation quarks dominate. We can therefore successfully reproduce the impact of triangle corrections by referring to the more succinct 
\bea \label{simTRI}
& \quad \quad \quad \quad \Delta r^{\text{triangles}}_{LQ} = T \left(m_b,m_t\right) \times \frac{\mathcal Y^2}{ M_{\mathcal S}^2} = \\
&  \frac{3}{64 \pi^2} \frac{m_t^4\left[1 + \log (m_t^4/M_{S_1}^4) \right] - m_b^4 \left[ 1 + \log (m_b^4/M_{S_1}^4)\right]}{{\left(m_t^2-m_b^2\right)}}  \\ & \quad \quad \quad \quad \times  \frac{\mathcal Y^2}{ M_{\mathcal S}^2} \,\,\,\,\,\, . 
\eea
Again, we have introduced the shortcut $T \left(m_b,m_t\right)$ for the recurring loop function and used the form $\mathcal Y^2$ to symbolically refer to quadratic combinations of the LQ Yukawas. In terms of these simple abbreviations, we can cover the dominant non-universal corrections to $\Delta r$ for all the five scalar LQs. 

\subsubsection*{$SU(2)_{EW}$ singlet}
%
\begin{figure}[h]
\includegraphics[scale=0.3]{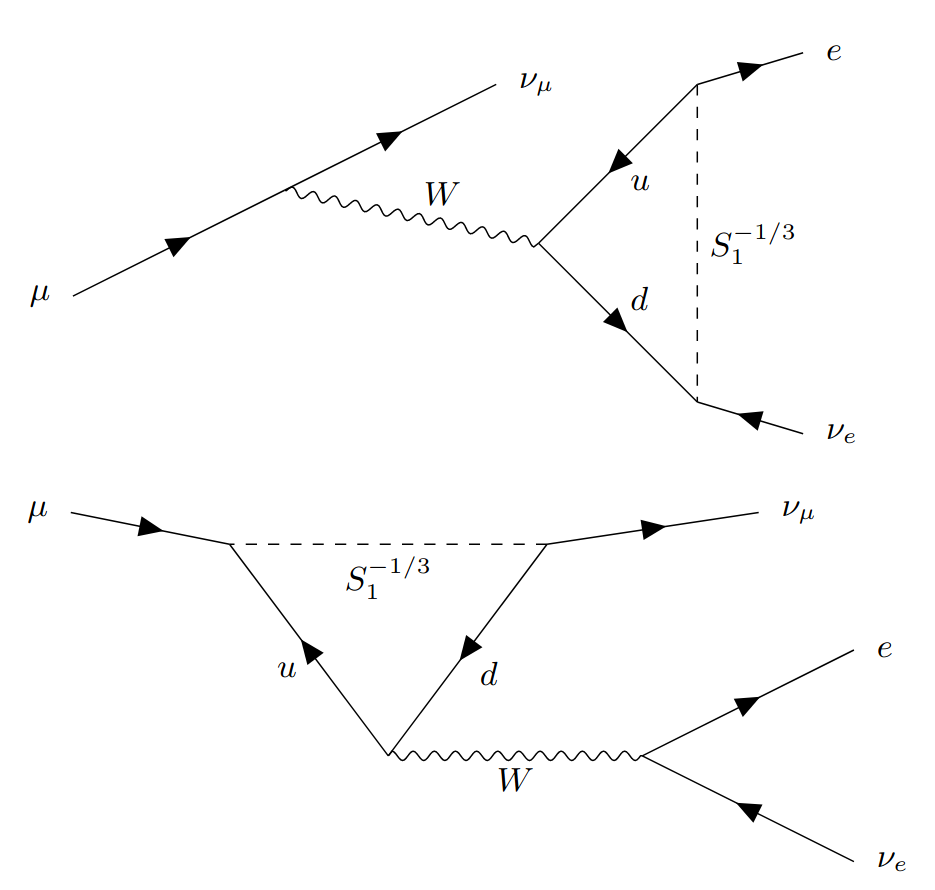}
\caption{
\label{fig:tri1S1} Triangle diagrams with a potential non-universal contribution to $\Delta r_{S_1}$.}
\end{figure}
\begin{figure}[h]
\includegraphics[scale=0.3]{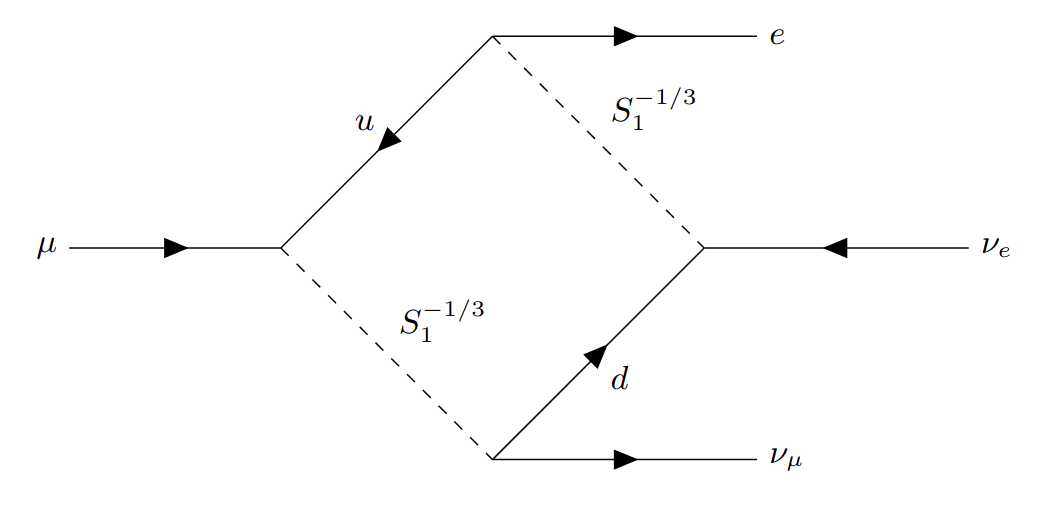}
\caption{\label{fig:box1S1}The unique box-type diagram with a potential non-universal contribution to $\Delta r_{S_1}$.}
\end{figure}
While $\tilde S_1 $ is relegated to irrelevance by the smallness of the lepton masses, the $S_1$ LQ can play a more impactful role. Being a $SU(2)_{EW}$ singlet, no corrections to the $W$ polarization are generated. Instead, we have purely non-universal corrections to $W \bar l l$ vertices \ref{fig:tri1S1} and box diagrams \ref{fig:box1S1}, not suppressed by any further chirality-flipping mass insertions. 
With the terminology introduced we have
\bea \label{S1triangles}
& \Delta r^{\text{triangles}}_{S_1} =  T\left(m_b, m_t\right) \times \frac{\mathcal Y_{LL}^{3,1} \,{}^2  + \mathcal Y_{LL}^{3,2} \,{}^2 }{M_{S_1}^2} \, ,
\eea
plus the box counterpart   
\bea \label{S1boxes}
& \Delta r^{\text{boxes}}_{S_1} = B\left(M_W,M_Z\right)  \times  \frac{\left(\sum_{i} \mathcal Y_{LL}^{i,1} \mathcal Y_{LL}^{i,2} \right)^2}{M_{S_1}^2} \, \quad .
\eea

\subsubsection*{$SU(2)_{EW}$ doublet}
 
\begin{figure}[h]
\includegraphics[scale=0.3]{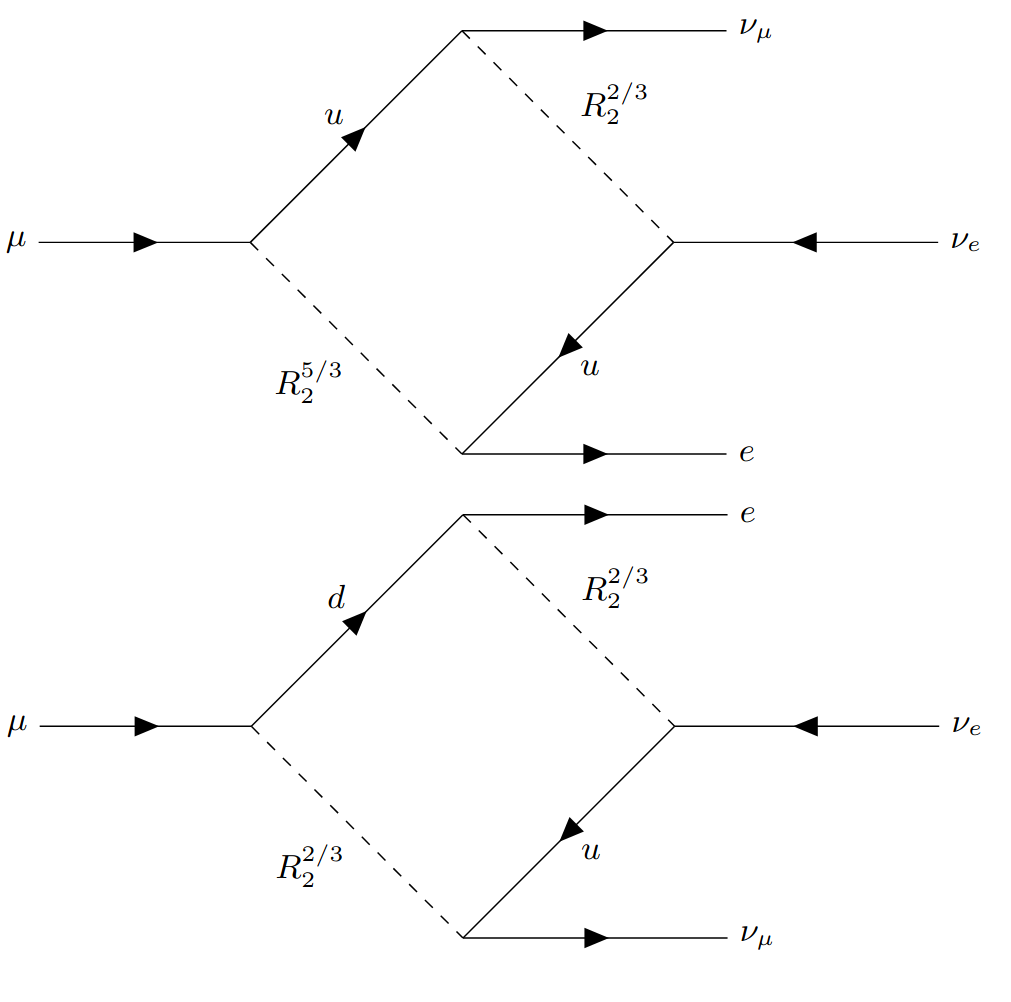}
\caption{\label{fig:box1R2} The diagrams providing the dominant non-universal contribution to $\Delta r$ for a $R_2$ LQ.}
\end{figure}

\begin{figure}[h]
\includegraphics[scale=0.3]{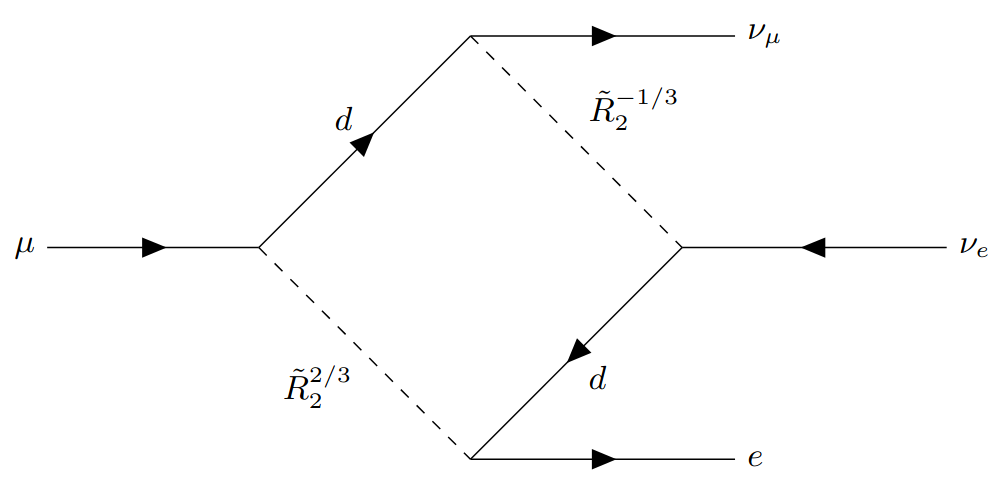}
\caption{\label{fig:box1R2t} The diagram providing the dominant non-universal contribution to $\Delta r$ for a $\tilde R_2$ LQ}
\end{figure}

When moving away from the simplest scenario of an $SU(2)_{EW}$ singlet, a richer selection of new sources for $\Delta r$ must be considered. This is generated by the added interaction with the $W$-boson as well as the presence and joint propagation of multiple new states. 
For both of the doublet LQs introduced in \ref{LQsDoubletF}, we have the complete selections of diagrammatic contributions appearing in \ref{deltar}. 
As expected at one-loop order, $\Delta r^{\text{self-energies}}$ can only give a universal correction and their role is therefore already included in the oblique parametrization of NP effects. We will not consider it here. 
For $\Delta r^{\text{triangles}}$ we have potentially many diagrams proportional to the LQ Yukawa couplings. We find, through a direct computation, their value is proportional to the light lepton masses, and therefore negligible. 
The leading non-oblique term for the doublet LQ comes from box diagrams. In the limit in which we can neglect powers of $m_t/\mathcal{M}_{LQ}$ we can adopt the reduction \ref{simpBOX} and write, for $R_2$ and \ref{fig:box1R2}, the familiar
\bea \label{deltarR2}
& \Delta r^{\text{boxes}}_{R_2} = B\left(M_W,M_Z\right)  \times  \frac{\sum_{i} (\mathcal Y_{RL}^{i,1})^2  \times \sum_{j} (\mathcal Y_{RL}^{i,2})^2 }{M_{R_2}^2} \, \quad .
\eea
At the same order ($\mathcal M_{LQ} > m_t$), we also find that the previous formula account for the box diagram propagating the content of the $\tilde R_2$ doublet \ref{fig:box1R2t}. We can therefore assign to $\Delta r^{\text{boxes}}_{\tilde R_2}$ the previous form  \ref{deltarR2} upon an obvious change of labels (the different Yukawa have already conveniently been represented with the same symbol $\mathcal Y_{RL}$ in \ref{LQsDoubletF}.


\subsubsection*{$SU(2)_{EW}$ triplet}

\begin{figure}[h]
\includegraphics[scale=0.3]{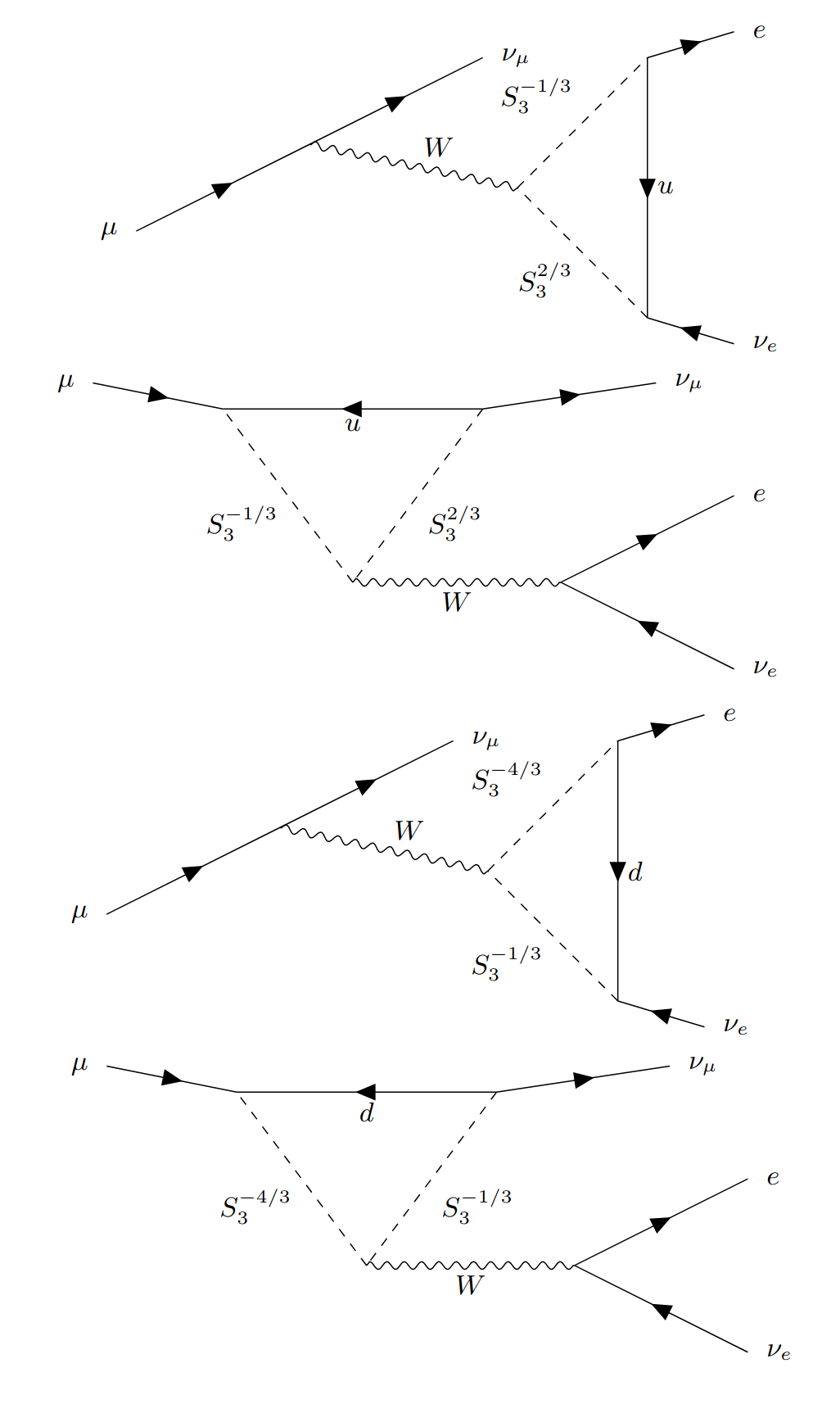}
\caption{\label{fig:tri1S3} Triangle diagrams with a non-universal contribution to $\Delta r_{S_3}$. Not included are those with only $S_3^{-1/3}$ propagating, equivalent to \ref{fig:tri1S1}.}
\end{figure}
\begin{figure}[h]
\includegraphics[scale=0.3]{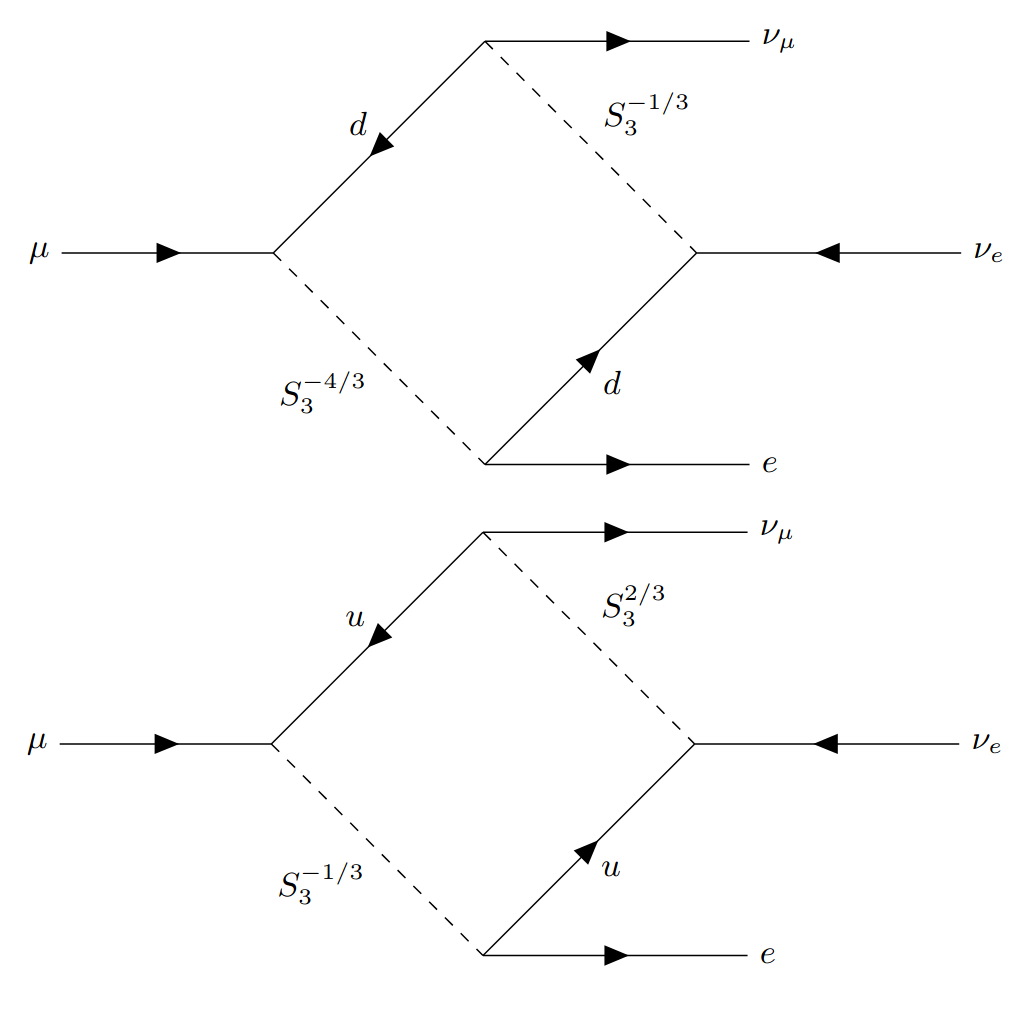}
\caption{\label{fig:box1S3} Box diagrams with a non-universal contribution to $\Delta r_{S_3}$. Not included are those with only $S_3^{-1/3}$ propagating, equivalent to \ref{fig:box1S1}.}
\end{figure}

The three charged scalars of the $S_3$ LQ represent a further step towards complexity and generate an increased number of diagrams. Again, we neglect the known oblique contribution of the polarization diagrams. We find that triangles with $S_3$ states can substantially shift $\Delta r$. In this set, those exclusively propagating the state $S_3^{-1/3}$ have an identical diagrammatic representation as those for the $S_1$ case \ref{fig:tri1S1}. The new ones are depicted in \ref{fig:tri1S3}. The collective value is  
\bea \label{S3triangles}
& \Delta r^{\text{triangles}}_{S_3}  =  - T\left(m_b, m_t\right) \times 
\frac{\left( \mathcal Y_{LL}^{3,1} \right)^2  + \left(\mathcal Y_{LL}^{3,2} \right)^2}{M_{S_3}^2} \, .
\eea
Finally, of the three box diagrams we found, one is accounted for \ref{fig:box1S1} and two new involve $S_3^{-4/3}$ and $S_3^{2/3}$ states \ref{fig:box1S3}. Together they generate a further non-universal addition to $\Delta r_{S_3}$
\bea \label{S3boxes}
& \Delta r^{\text{boxes}}_{S_3} = B\left(M_W,M_Z\right)  \times \\
& \quad \quad \quad \times \frac{\left(\sum_{i} \mathcal Y_{LL}^{i,1} \mathcal Y_{LL}^{i,2} \right)^2 + 4 \sum_{i} \left(\mathcal Y_{LL}^{i,1} \right)^2 \sum_{j} \left(\mathcal Y_{LL}^{j,2} \right)^2}{M_{S_3}^2} \,  .
\eea

\section{Results}
\begin{figure}[h]
\includegraphics[scale=0.5]{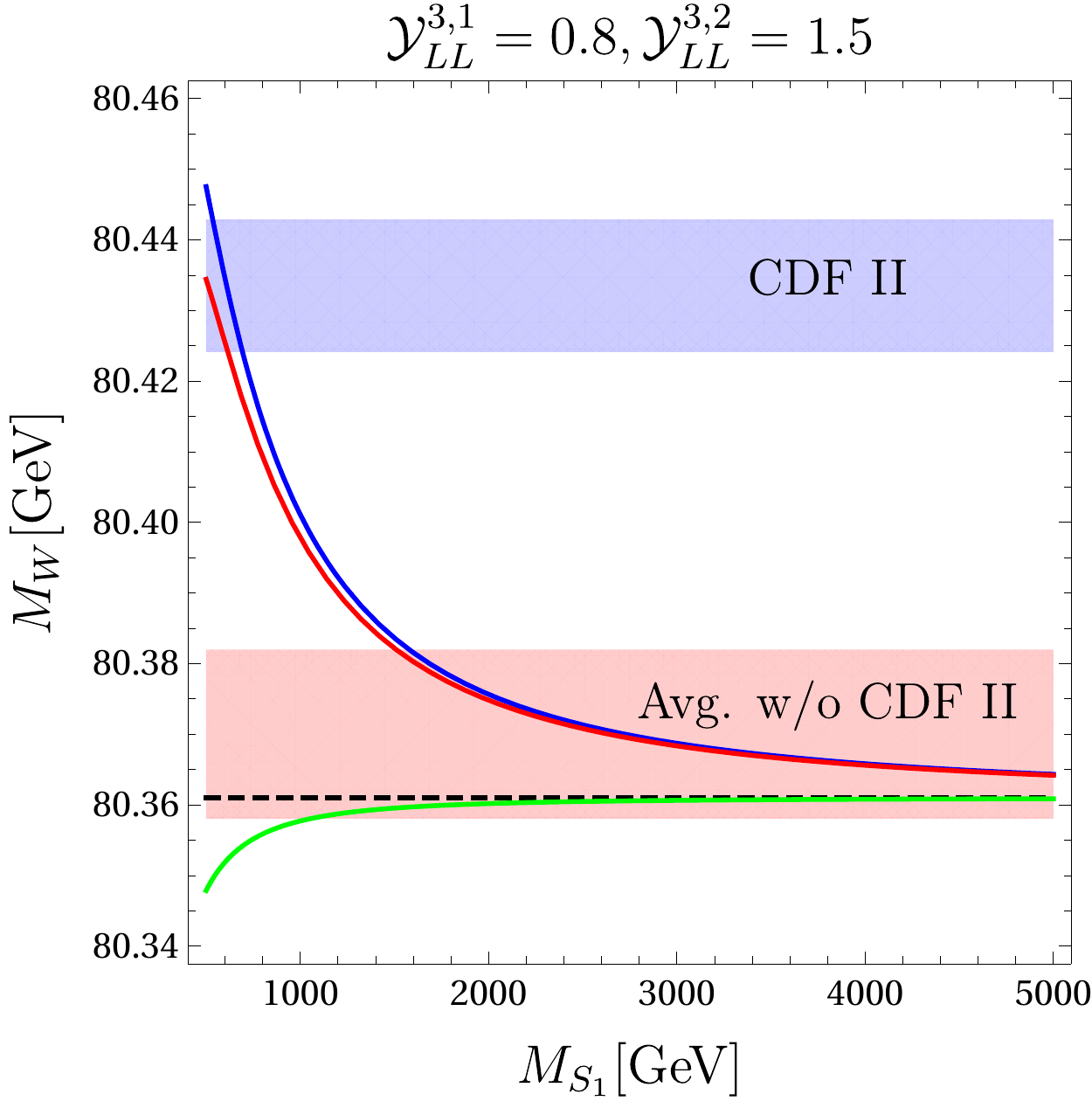}
\includegraphics[scale=0.5]{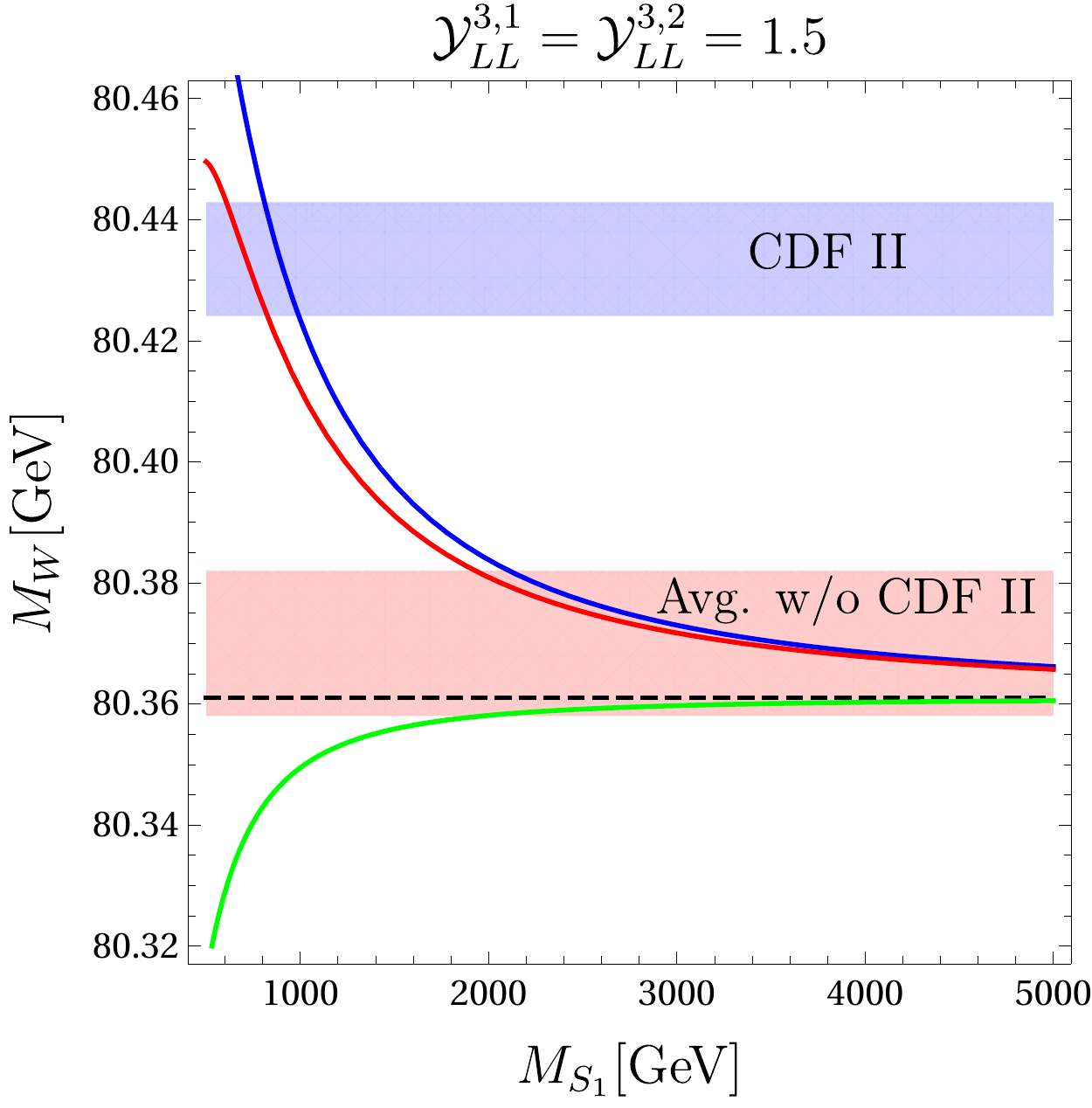}
\caption{For the $S_1$ scenario: relevance of triangle (blue line), box (green line) and the overall contribution (red line) to $M_W$ with only two couplings $\mathcal{Y}_{LL}^{3,1}$ and $\mathcal{Y}_{LL}^{3,2}$ non zero. Dashed black line is the benchmark value $M_W^{2013}$.}
\label{fig:S1plot23}
\end{figure}

\begin{figure}[h]
\includegraphics[scale=0.5]{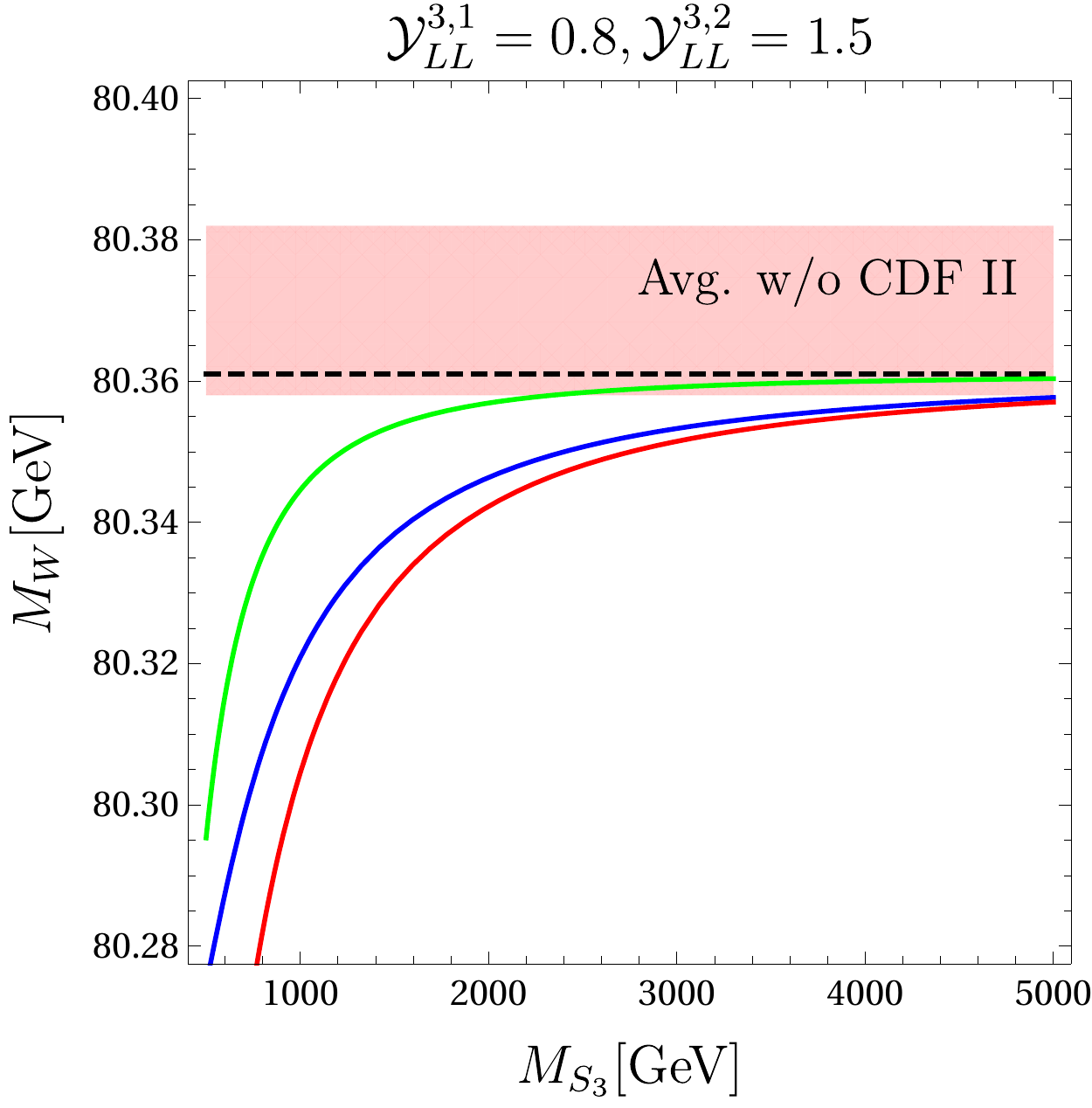}
\includegraphics[scale=0.5]{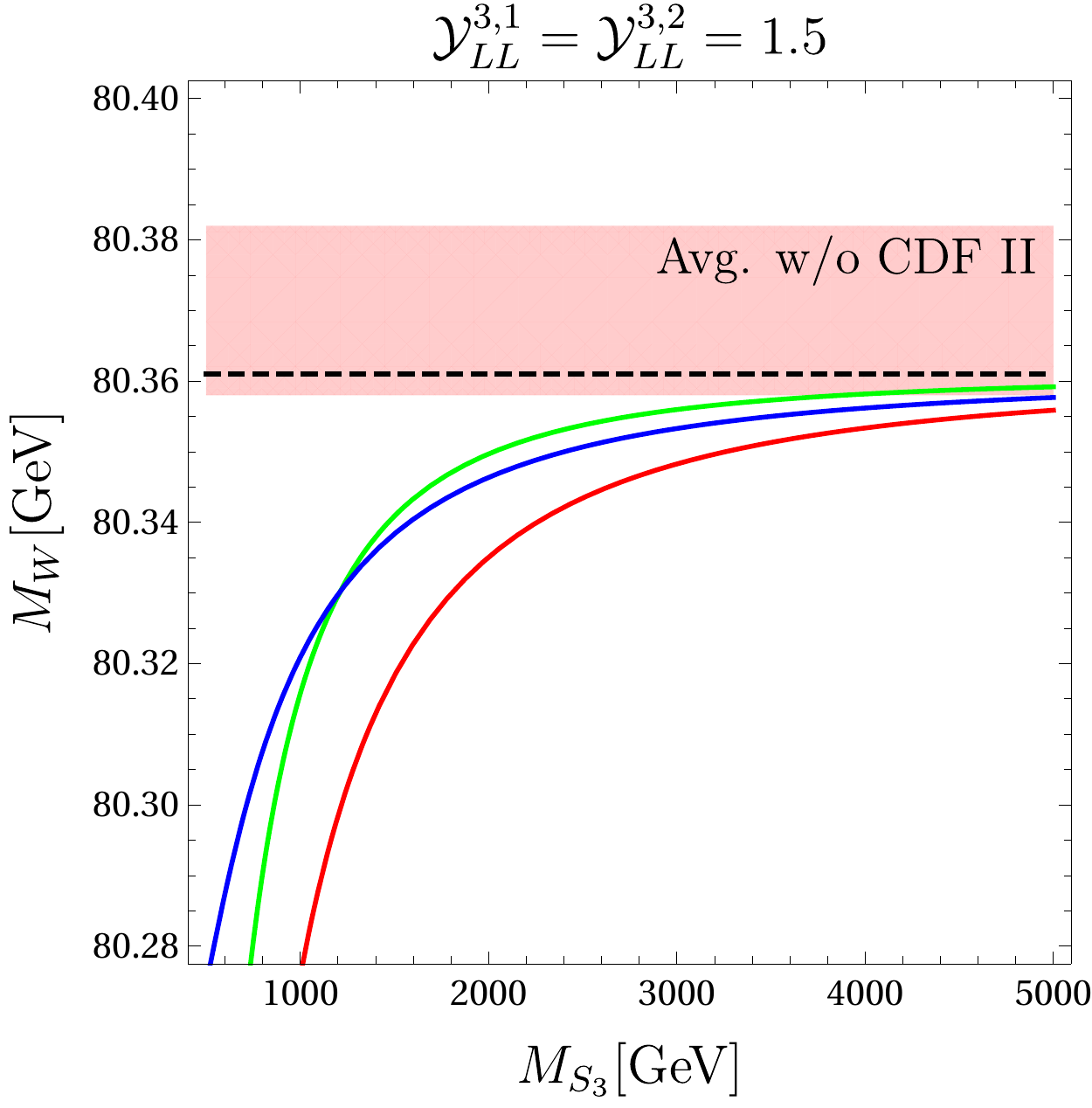}
\caption{For the $S_3$ scenario: relevance of triangle (blue line), box (green line) and the overall contribution (red line) to $M_W$ with only the two couplings $\mathcal{Y}_{LL}^{3,1}$ and $\mathcal{Y}_{LL}^{3,2}$ non zero. Dashed black line is the benchmark value $M_W^{2013}$.}
\label{fig:S3plot23}
\end{figure}

\begin{figure}
\includegraphics[scale=0.5]{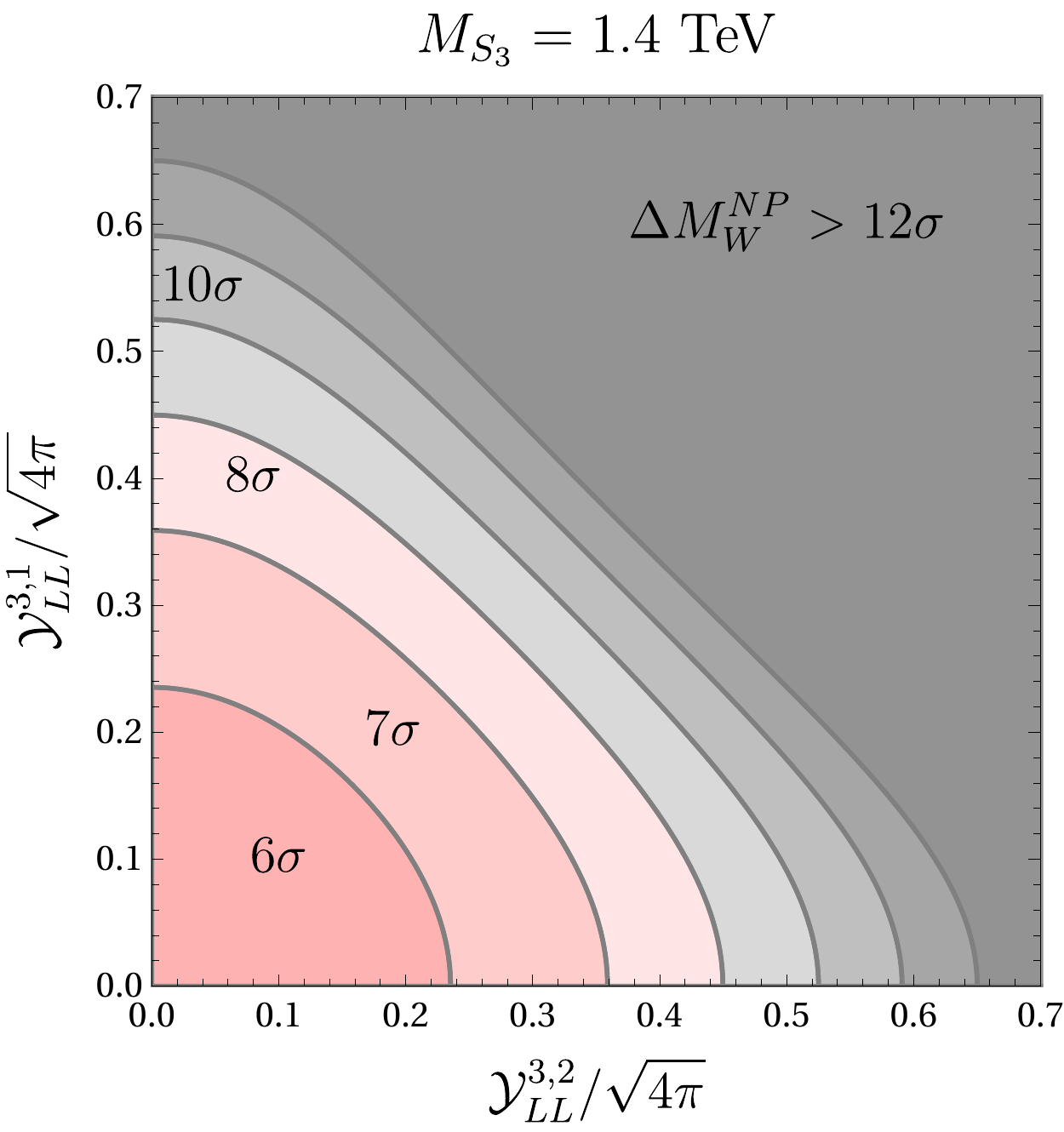}
\includegraphics[scale=0.5]{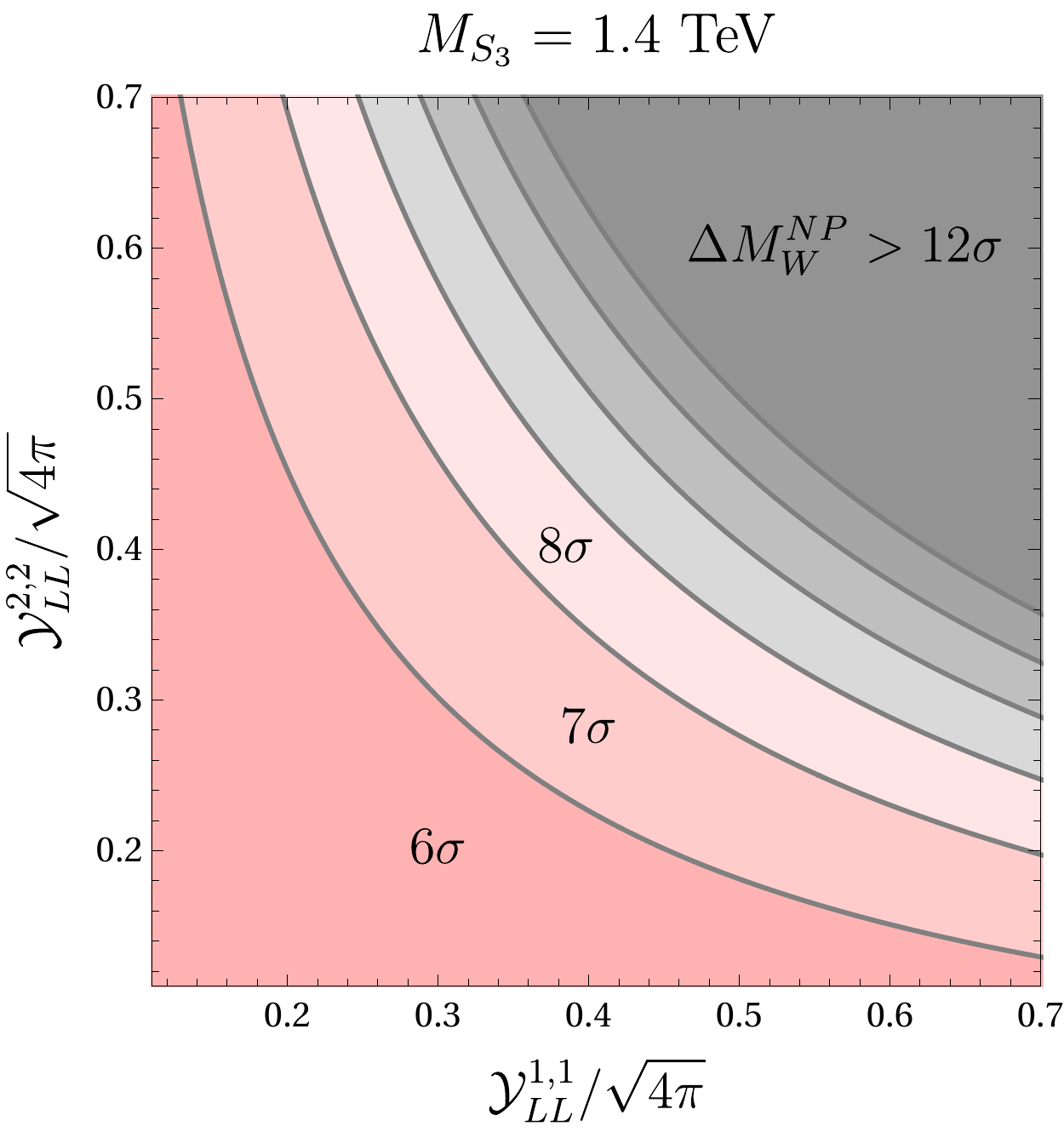}
\caption{The regions in the parameter space of $S_3$ with sizable contributions to $\Delta r^{NP}$. Shades from red to grey separate regions with a minimum value of $6\sigma$ to values greater than $12\sigma$ from $M_W^{\text{CDF II}}$.
}
\label{fig:S3plot23x}
\end{figure}

Experimental determinations of $M_W$(\cite{LHCb:2021bjt} being the only exception) have consistently pointed towards values larger than the corresponding SM prediction. Until recently, daring NP interpretations of this pattern were restrained by the mismatch between the small theoretical error and the bigger one assigned to the experimental measurement.
It is therefore an intriguing result that the most recent of these determinations, carried by the CDF collaboration with a data set of 8.8 fb${}^{-1}$ and a sensibly increased precision, supports the same deviation and raises the tension, if taken individually, up to $7\sigma$. 
\begin{figure}[h]
\includegraphics[scale=0.6]{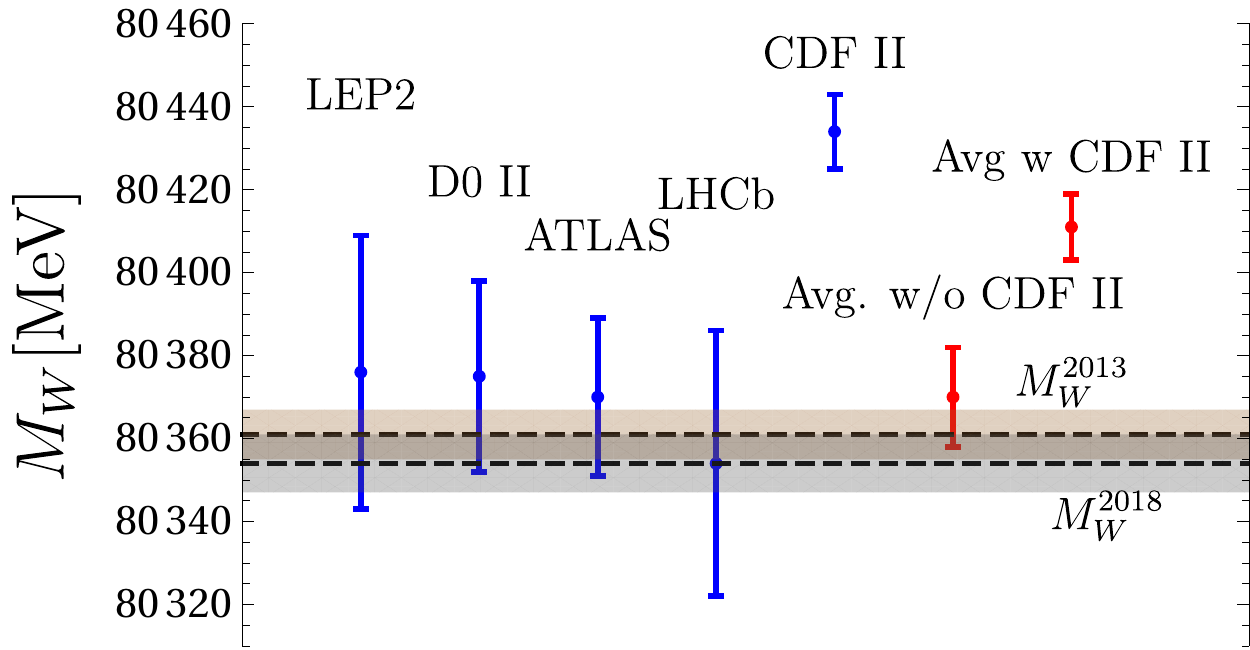}
\caption{Status of W mass measurements (blue lines) from \cite{ALEPH:2005ab,D0:2012kms,ATLAS:2017rzl,LHCb:2021bjt} and CDF II \cite{CDFII}. Red lines indicate fits of \cite{Bagnaschi:2022whn} describing the experimental averages with and without CDF II. Lastly, the horizontal bands correspond to the SM predictions of \cite{Heinemeyer:2013dia,Haller:2018nnx}. }
\label{fig:dataMW}
\end{figure}
We can entertain the idea that this deviation should be explained by NP. Nevertheless, this is not the unique motivation behind our work. Indeed the last CDF measurement is one among different efforts of the experimental community to raise the status of $M_W$ to a precision observable. A future, promising one, against which we ultimately weigh the relevance of NP-induced $\Delta M_W$ \ref{MWNP}, is based on a projected $12$ ab$^{-1}$ of data at the Future Circular Collider (FCC-ee). Studies of $e^+ e^- \rightarrow W^+ W^-$ around the boson pair threshold are realistically expected to bring the uncertainty down to the $0.5$ MeV level \cite{FCC:2018byv} and a parallel theoretical effort in the computation of the $\Delta r$ is therefore mandatory.
We now summarize the main results obtained through the simplified, but accurate formulas presented in this work, accounting for the leading non-oblique contributions to $M_W$ for the case of scalar LQs. In linking $\Delta r^{NP}$ to the value of $M_W$ in \ref{MWNP} we arbitrarily adopt $M_W^{\text{2013}}$ of \cite{Heinemeyer:2013dia} as representative of the (dominant) SM prediction. 
Different Yukawa couplings are involved in the triangle \ref{S1triangles} and box \ref{S1boxes} contributions to $M_W$ in the LQ $S_1$ case. A comparative study of the relevance of the different diagrams points to a general pattern of large positive contributions from the triangles and a small negative one from the boxes \ref{fig:S1plot23}. 
Therefore, the selected couplings that can significantly contribute to $M_W$ are $\mathcal Y_{LL}^{3,1}$ and $\mathcal Y_{LL}^{3,2}$. This must not be interpreted as a call for the remaining Yukawa couplings to be zero. Indeed they might play an important role in preserving the viability of the parameter space by affecting further observables, thus avoiding tight phenomenological constraints. For instance, we can ignore, at first look, the  bounds coming from measurements of the anomalous magnetic moment of light charged leptons, which would require, on top of $\mathcal Y_{LL}^{3,1}$ and $\mathcal Y_{LL}^{3,2}$, a non-zero textures for $\mathcal{Y}_{RR}$.
Similarly, we can optimistically assume, for sake of illustration, that extra couplings affect the decay branching ratios of the LQs, which shape the assumptions behind the rigid collider bounds in the $\mathcal{M}_{S}\sim 0.8 - 1.5$ TeV range. In this way, we might hope of eluding the simplified leptoquark scenarios analyzed in \cite{CMS:2018txo,ATLAS:2020dsk,CMS:2018oaj,ATLAS:2019qpq,ATLAS:2020xov,ATLAS:2021oiz}, circumventing, for instance, relations as $Br\left(S_{1} \rightarrow t \bar \tau \right) = 1 - Br\left(S_{1} \rightarrow b \bar \nu_{\tau} \right)$ or other branching-related assumptions. 
\begin{figure}[h]
\vspace{2ex}%
\includegraphics[scale=0.5]{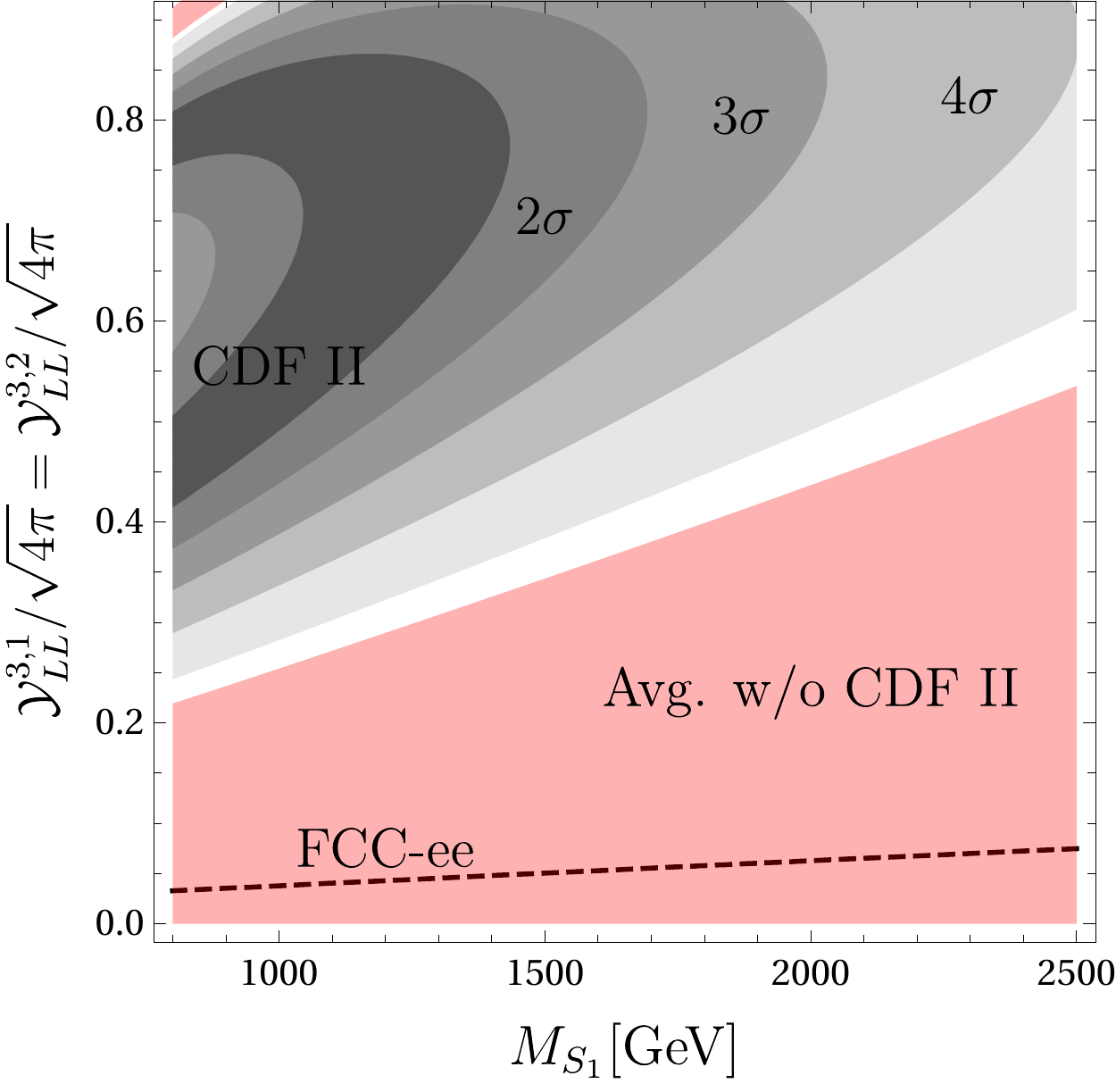}
\caption{The regions in the parameter space of $S_1$ with sizable contributions to $\Delta r^{NP} \sim \Delta r^{\text{triangles}}_{S_1}$. The Yukawa sector is probed through the variable $\xi = \mathcal{Y}_{LL}^{3,1} / \sqrt{4\pi} = \mathcal{Y}_{LL}^{2,1} / \sqrt{4\pi}$. Shades of grey point to regions in the $1\sigma-5\sigma$ interval from the new CDF $M_W$ measure. Light red area complies with the $1\sigma$ region of the global average estimate of $M_W$ without the last CDF value \cite{Bagnaschi:2022whn}. The region above the black dashed line is under potential future FCC detection \cite{FCC:2018byv}.}
\label{fig:S1plot1}
\end{figure}
We want to draw here the main effects that scalar LQs generate on $\Delta r$, and  considerations about the viability of the parameter space are postponed to a more detailed exploration of selected models including the LQ mechanism. In fig.~\ref{fig:S1plot1} we show that regions exist, in the $M_{S_1}-\mathcal Y_{LL}$ plane, that can ease the distance with the CDF II measure. We notice ample, perturbative areas that can successfully reduce the current tension and, in a more conservative approach, provide support to oblique corrections along the same direction. 
The patterns provided by the other LQs are clearly distinguishable from $S_1$. Considering $S_3$, we see that boxes and triangles collectively worsen the separation with the measured value \ref{fig:S3plot23}. We illustrate in fig.~\ref{fig:S3plot23x} how, even for masses up to $1.4$ TeV, perturbative regions in the space of Yukawa couplings would raise the discrepancy towards unacceptable values. 
Finally, while the Yukawa structures of the box contribution are in general different between $S_1$ and the doublet LQ models, they converge to the same formula plotted in \ref{fig:S1plot23} upon the identification $\mathcal{Y}_{LL}^{3,1} = \mathcal{Y}_{RL}^{3,1} $ and $\mathcal{Y}_{LL}^{3,2} = \mathcal{Y}_{RL}^{3,2} $
and setting the remaining couplings to zero. We can therefore draw a similar conclusion as for the $S_3$ scenario that the present tension in $M_W$ is worsened by flavour-dependent interactions with the doublet scalar LQs.
%


\section{Conclusions}
We have computed the leading non-universal corrections to the W-boson mass for the five scalar LQs allowed by the SM gauge symmetries. We provided simple formulas to account for the effect of Yukawa-type interactions between the SM matter content and the LQs. We found that the $S_1$ LQ can alleviate the current tension while $S_3$, $R_2$ and $\tilde R_2$ can exacerbate it. Aside from $\tilde S_1$, all the presented non-universal signatures can be detected by future FCC-ee probing.
\\\\
\noindent \textbf{Acknowledgement.} 
We thank Niko Koivunen, Luca Marzola, Aurora Melis and  Marco Piva for clarifications and relevant comments. 
Feynman diagrams images have been generated via \cite{Ellis:2016jkw}.
This work was supported by the Estonian Research Council grant PRG1677.

\newpage

\bibliography{main}

\begin{thebibliography}{126}
\expandafter\ifx\csname natexlab\endcsname\relax\def\natexlab#1{#1}\fi
\expandafter\ifx\csname bibnamefont\endcsname\relax
  \def\bibnamefont#1{#1}\fi
\expandafter\ifx\csname bibfnamefont\endcsname\relax
  \def\bibfnamefont#1{#1}\fi
\expandafter\ifx\csname citenamefont\endcsname\relax
  \def\citenamefont#1{#1}\fi
\expandafter\ifx\csname url\endcsname\relax
  \def\url#1{\texttt{#1}}\fi
\expandafter\ifx\csname urlprefix\endcsname\relax\def\urlprefix{URL }\fi
\providecommand{\bibinfo}[2]{#2}
\providecommand{\eprint}[2][]{\url{#2}}

\bibitem[{\citenamefont{Veltman}(1977)}]{Veltman:1977kh}
\bibinfo{author}{\bibfnamefont{M.~J.~G.} \bibnamefont{Veltman}},
  \bibinfo{journal}{Nucl. Phys. B} \textbf{\bibinfo{volume}{123}},
  \bibinfo{pages}{89} (\bibinfo{year}{1977}).

\bibitem[{\citenamefont{Abi et~al.}(2021)}]{Muong-2:2021ojo}
\bibinfo{author}{\bibfnamefont{B.}~\bibnamefont{Abi}} \bibnamefont{et~al.}
  (\bibinfo{collaboration}{Muon g-2}), \bibinfo{journal}{Phys. Rev. Lett.}
  \textbf{\bibinfo{volume}{126}}, \bibinfo{pages}{141801}
  (\bibinfo{year}{2021}), \eprint{2104.03281}.

\bibitem[{\citenamefont{Bennett et~al.}(2006)}]{Muong-2:2006rrc}
\bibinfo{author}{\bibfnamefont{G.~W.} \bibnamefont{Bennett}}
  \bibnamefont{et~al.} (\bibinfo{collaboration}{Muon g-2}),
  \bibinfo{journal}{Phys. Rev. D} \textbf{\bibinfo{volume}{73}},
  \bibinfo{pages}{072003} (\bibinfo{year}{2006}), \eprint{hep-ex/0602035}.

\bibitem[{\citenamefont{Aoyama et~al.}(2020)}]{Aoyama:2020ynm}
\bibinfo{author}{\bibfnamefont{T.}~\bibnamefont{Aoyama}} \bibnamefont{et~al.},
  \bibinfo{journal}{Phys. Rept.} \textbf{\bibinfo{volume}{887}},
  \bibinfo{pages}{1} (\bibinfo{year}{2020}), \eprint{2006.04822}.

\bibitem[{\citenamefont{Czarnecki et~al.}(2003)\citenamefont{Czarnecki,
  Marciano, and Vainshtein}}]{Czarnecki:2002nt}
\bibinfo{author}{\bibfnamefont{A.}~\bibnamefont{Czarnecki}},
  \bibinfo{author}{\bibfnamefont{W.~J.} \bibnamefont{Marciano}},
  \bibnamefont{and}
  \bibinfo{author}{\bibfnamefont{A.}~\bibnamefont{Vainshtein}},
  \bibinfo{journal}{Phys. Rev. D} \textbf{\bibinfo{volume}{67}},
  \bibinfo{pages}{073006} (\bibinfo{year}{2003}), \bibinfo{note}{[Erratum:
  Phys.Rev.D 73, 119901 (2006)]}, \eprint{hep-ph/0212229}.

\bibitem[{\citenamefont{Gnendiger et~al.}(2013)\citenamefont{Gnendiger,
  St\"ockinger, and St\"ockinger-Kim}}]{Gnendiger:2013pva}
\bibinfo{author}{\bibfnamefont{C.}~\bibnamefont{Gnendiger}},
  \bibinfo{author}{\bibfnamefont{D.}~\bibnamefont{St\"ockinger}},
  \bibnamefont{and}
  \bibinfo{author}{\bibfnamefont{H.}~\bibnamefont{St\"ockinger-Kim}},
  \bibinfo{journal}{Phys. Rev. D} \textbf{\bibinfo{volume}{88}},
  \bibinfo{pages}{053005} (\bibinfo{year}{2013}), \eprint{1306.5546}.

\bibitem[{\citenamefont{Keshavarzi et~al.}(2018)\citenamefont{Keshavarzi,
  Nomura, and Teubner}}]{Keshavarzi:2018mgv}
\bibinfo{author}{\bibfnamefont{A.}~\bibnamefont{Keshavarzi}},
  \bibinfo{author}{\bibfnamefont{D.}~\bibnamefont{Nomura}}, \bibnamefont{and}
  \bibinfo{author}{\bibfnamefont{T.}~\bibnamefont{Teubner}},
  \bibinfo{journal}{Phys. Rev. D} \textbf{\bibinfo{volume}{97}},
  \bibinfo{pages}{114025} (\bibinfo{year}{2018}), \eprint{1802.02995}.

\bibitem[{\citenamefont{Colangelo et~al.}(2019)\citenamefont{Colangelo,
  Hoferichter, and Stoffer}}]{Colangelo:2018mtw}
\bibinfo{author}{\bibfnamefont{G.}~\bibnamefont{Colangelo}},
  \bibinfo{author}{\bibfnamefont{M.}~\bibnamefont{Hoferichter}},
  \bibnamefont{and} \bibinfo{author}{\bibfnamefont{P.}~\bibnamefont{Stoffer}},
  \bibinfo{journal}{JHEP} \textbf{\bibinfo{volume}{02}}, \bibinfo{pages}{006}
  (\bibinfo{year}{2019}), \eprint{1810.00007}.

\bibitem[{\citenamefont{Kurz et~al.}(2014)\citenamefont{Kurz, Liu, Marquard,
  and Steinhauser}}]{Kurz:2014wya}
\bibinfo{author}{\bibfnamefont{A.}~\bibnamefont{Kurz}},
  \bibinfo{author}{\bibfnamefont{T.}~\bibnamefont{Liu}},
  \bibinfo{author}{\bibfnamefont{P.}~\bibnamefont{Marquard}}, \bibnamefont{and}
  \bibinfo{author}{\bibfnamefont{M.}~\bibnamefont{Steinhauser}},
  \bibinfo{journal}{Phys. Lett. B} \textbf{\bibinfo{volume}{734}},
  \bibinfo{pages}{144} (\bibinfo{year}{2014}), \eprint{1403.6400}.

\bibitem[{\citenamefont{Melnikov and Vainshtein}(2004)}]{Melnikov:2003xd}
\bibinfo{author}{\bibfnamefont{K.}~\bibnamefont{Melnikov}} \bibnamefont{and}
  \bibinfo{author}{\bibfnamefont{A.}~\bibnamefont{Vainshtein}},
  \bibinfo{journal}{Phys. Rev. D} \textbf{\bibinfo{volume}{70}},
  \bibinfo{pages}{113006} (\bibinfo{year}{2004}), \eprint{hep-ph/0312226}.

\bibitem[{\citenamefont{Athron et~al.}(2021)\citenamefont{Athron, Bal\'azs,
  Jacob, Kotlarski, St\"ockinger, and St\"ockinger-Kim}}]{Athron:2021iuf}
\bibinfo{author}{\bibfnamefont{P.}~\bibnamefont{Athron}},
  \bibinfo{author}{\bibfnamefont{C.}~\bibnamefont{Bal\'azs}},
  \bibinfo{author}{\bibfnamefont{D.~H.~J.} \bibnamefont{Jacob}},
  \bibinfo{author}{\bibfnamefont{W.}~\bibnamefont{Kotlarski}},
  \bibinfo{author}{\bibfnamefont{D.}~\bibnamefont{St\"ockinger}},
  \bibnamefont{and}
  \bibinfo{author}{\bibfnamefont{H.}~\bibnamefont{St\"ockinger-Kim}},
  \bibinfo{journal}{JHEP} \textbf{\bibinfo{volume}{09}}, \bibinfo{pages}{080}
  (\bibinfo{year}{2021}), \eprint{2104.03691}.

\bibitem[{\citenamefont{Aaltonen et~al.}(2022)}]{CDFII}
\bibinfo{author}{\bibfnamefont{T.}~\bibnamefont{Aaltonen}}
  \bibnamefont{et~al.}, \bibinfo{journal}{Science}
  \textbf{\bibinfo{volume}{376}}, \bibinfo{pages}{170} (\bibinfo{year}{2022}),
  \eprint{https://www.science.org/doi/pdf/10.1126/science.abk1781},
  \urlprefix\url{https://www.science.org/doi/abs/10.1126/science.abk1781}.

\bibitem[{\citenamefont{Haller et~al.}(2018)\citenamefont{Haller, Hoecker,
  Kogler, M\"onig, Peiffer, and Stelzer}}]{Haller:2018nnx}
\bibinfo{author}{\bibfnamefont{J.}~\bibnamefont{Haller}},
  \bibinfo{author}{\bibfnamefont{A.}~\bibnamefont{Hoecker}},
  \bibinfo{author}{\bibfnamefont{R.}~\bibnamefont{Kogler}},
  \bibinfo{author}{\bibfnamefont{K.}~\bibnamefont{M\"onig}},
  \bibinfo{author}{\bibfnamefont{T.}~\bibnamefont{Peiffer}}, \bibnamefont{and}
  \bibinfo{author}{\bibfnamefont{J.}~\bibnamefont{Stelzer}},
  \bibinfo{journal}{Eur. Phys. J. C} \textbf{\bibinfo{volume}{78}},
  \bibinfo{pages}{675} (\bibinfo{year}{2018}), \eprint{1803.01853}.

\bibitem[{\citenamefont{Abada et~al.}(2019)}]{FCC:2018byv}
\bibinfo{author}{\bibfnamefont{A.}~\bibnamefont{Abada}} \bibnamefont{et~al.}
  (\bibinfo{collaboration}{FCC}), \bibinfo{journal}{Eur. Phys. J. C}
  \textbf{\bibinfo{volume}{79}}, \bibinfo{pages}{474} (\bibinfo{year}{2019}).

\bibitem[{\citenamefont{Strumia}(2022)}]{Strumia:2022qkt}
\bibinfo{author}{\bibfnamefont{A.}~\bibnamefont{Strumia}},
  \bibinfo{journal}{JHEP} \textbf{\bibinfo{volume}{08}}, \bibinfo{pages}{248}
  (\bibinfo{year}{2022}), \eprint{2204.04191}.

\bibitem[{\citenamefont{Asadi et~al.}(2022)\citenamefont{Asadi, Cesarotti,
  Fraser, Homiller, and Parikh}}]{Asadi:2022xiy}
\bibinfo{author}{\bibfnamefont{P.}~\bibnamefont{Asadi}},
  \bibinfo{author}{\bibfnamefont{C.}~\bibnamefont{Cesarotti}},
  \bibinfo{author}{\bibfnamefont{K.}~\bibnamefont{Fraser}},
  \bibinfo{author}{\bibfnamefont{S.}~\bibnamefont{Homiller}}, \bibnamefont{and}
  \bibinfo{author}{\bibfnamefont{A.}~\bibnamefont{Parikh}}
  (\bibinfo{year}{2022}), \eprint{2204.05283}.

\bibitem[{\citenamefont{Primulando et~al.}(2022)\citenamefont{Primulando,
  Julio, and Uttayarat}}]{Primulando:2022vip}
\bibinfo{author}{\bibfnamefont{R.}~\bibnamefont{Primulando}},
  \bibinfo{author}{\bibfnamefont{J.}~\bibnamefont{Julio}}, \bibnamefont{and}
  \bibinfo{author}{\bibfnamefont{P.}~\bibnamefont{Uttayarat}}
  (\bibinfo{year}{2022}), \eprint{2211.16021}.

\bibitem[{\citenamefont{Bahl et~al.}(2022)\citenamefont{Bahl, Chiu, Gao, Wang,
  and Zhong}}]{Bahl:2022gqg}
\bibinfo{author}{\bibfnamefont{H.}~\bibnamefont{Bahl}},
  \bibinfo{author}{\bibfnamefont{W.~H.} \bibnamefont{Chiu}},
  \bibinfo{author}{\bibfnamefont{C.}~\bibnamefont{Gao}},
  \bibinfo{author}{\bibfnamefont{L.-T.} \bibnamefont{Wang}}, \bibnamefont{and}
  \bibinfo{author}{\bibfnamefont{Y.-M.} \bibnamefont{Zhong}},
  \bibinfo{journal}{Eur. Phys. J. C} \textbf{\bibinfo{volume}{82}},
  \bibinfo{pages}{944} (\bibinfo{year}{2022}), \eprint{2207.04059}.

\bibitem[{\citenamefont{Centelles~Chuli\'a
  et~al.}(2022)\citenamefont{Centelles~Chuli\'a, Srivastava, and
  Yadav}}]{CentellesChulia:2022vpz}
\bibinfo{author}{\bibfnamefont{S.}~\bibnamefont{Centelles~Chuli\'a}},
  \bibinfo{author}{\bibfnamefont{R.}~\bibnamefont{Srivastava}},
  \bibnamefont{and} \bibinfo{author}{\bibfnamefont{S.}~\bibnamefont{Yadav}}
  (\bibinfo{year}{2022}), \eprint{2206.11903}.

\bibitem[{\citenamefont{Kawamura and Raby}(2022)}]{Kawamura:2022fhm}
\bibinfo{author}{\bibfnamefont{J.}~\bibnamefont{Kawamura}} \bibnamefont{and}
  \bibinfo{author}{\bibfnamefont{S.}~\bibnamefont{Raby}},
  \bibinfo{journal}{Phys. Rev. D} \textbf{\bibinfo{volume}{106}},
  \bibinfo{pages}{035009} (\bibinfo{year}{2022}), \eprint{2205.10480}.

\bibitem[{\citenamefont{Chowdhury and Saad}(2022)}]{Chowdhury:2022dps}
\bibinfo{author}{\bibfnamefont{T.~A.} \bibnamefont{Chowdhury}}
  \bibnamefont{and} \bibinfo{author}{\bibfnamefont{S.}~\bibnamefont{Saad}},
  \bibinfo{journal}{Phys. Rev. D} \textbf{\bibinfo{volume}{106}},
  \bibinfo{pages}{055017} (\bibinfo{year}{2022}), \eprint{2205.03917}.

\bibitem[{\citenamefont{Isaacson et~al.}(2022)\citenamefont{Isaacson, Fu, and
  Yuan}}]{Isaacson:2022rts}
\bibinfo{author}{\bibfnamefont{J.}~\bibnamefont{Isaacson}},
  \bibinfo{author}{\bibfnamefont{Y.}~\bibnamefont{Fu}}, \bibnamefont{and}
  \bibinfo{author}{\bibfnamefont{C.~P.} \bibnamefont{Yuan}}
  (\bibinfo{year}{2022}), \eprint{2205.02788}.

\bibitem[{\citenamefont{He}(2022)}]{He:2022zjz}
\bibinfo{author}{\bibfnamefont{S.-P.} \bibnamefont{He}} (\bibinfo{year}{2022}),
  \eprint{2205.02088}.

\bibitem[{\citenamefont{Dcruz and Thapa}(2022)}]{Dcruz:2022dao}
\bibinfo{author}{\bibfnamefont{R.}~\bibnamefont{Dcruz}} \bibnamefont{and}
  \bibinfo{author}{\bibfnamefont{A.}~\bibnamefont{Thapa}}
  (\bibinfo{year}{2022}), \eprint{2205.02217}.

\bibitem[{\citenamefont{Kim}(2022)}]{Kim:2022xuo}
\bibinfo{author}{\bibfnamefont{J.}~\bibnamefont{Kim}}, \bibinfo{journal}{Phys.
  Lett. B} \textbf{\bibinfo{volume}{832}}, \bibinfo{pages}{137220}
  (\bibinfo{year}{2022}), \eprint{2205.01437}.

\bibitem[{\citenamefont{Barman et~al.}(2022)\citenamefont{Barman, Das, and
  Sengupta}}]{Barman:2022qix}
\bibinfo{author}{\bibfnamefont{B.}~\bibnamefont{Barman}},
  \bibinfo{author}{\bibfnamefont{A.}~\bibnamefont{Das}}, \bibnamefont{and}
  \bibinfo{author}{\bibfnamefont{S.}~\bibnamefont{Sengupta}}
  (\bibinfo{year}{2022}), \eprint{2205.01699}.

\bibitem[{\citenamefont{Wang et~al.}(2022)\citenamefont{Wang, Bi, Yin, and
  Yu}}]{Wang:2022dte}
\bibinfo{author}{\bibfnamefont{J.-W.} \bibnamefont{Wang}},
  \bibinfo{author}{\bibfnamefont{X.-J.} \bibnamefont{Bi}},
  \bibinfo{author}{\bibfnamefont{P.-F.} \bibnamefont{Yin}}, \bibnamefont{and}
  \bibinfo{author}{\bibfnamefont{Z.-H.} \bibnamefont{Yu}},
  \bibinfo{journal}{Phys. Rev. D} \textbf{\bibinfo{volume}{106}},
  \bibinfo{pages}{055001} (\bibinfo{year}{2022}), \eprint{2205.00783}.

\bibitem[{\citenamefont{Botella et~al.}(2022)\citenamefont{Botella,
  Cornet-Gomez, Mir\'o, and Nebot}}]{Botella:2022rte}
\bibinfo{author}{\bibfnamefont{F.~J.} \bibnamefont{Botella}},
  \bibinfo{author}{\bibfnamefont{F.}~\bibnamefont{Cornet-Gomez}},
  \bibinfo{author}{\bibfnamefont{C.}~\bibnamefont{Mir\'o}}, \bibnamefont{and}
  \bibinfo{author}{\bibfnamefont{M.}~\bibnamefont{Nebot}},
  \bibinfo{journal}{Eur. Phys. J. C} \textbf{\bibinfo{volume}{82}},
  \bibinfo{pages}{915} (\bibinfo{year}{2022}), \eprint{2205.01115}.

\bibitem[{\citenamefont{Gupta}(2022)}]{Gupta:2022lrt}
\bibinfo{author}{\bibfnamefont{R.~S.} \bibnamefont{Gupta}}
  (\bibinfo{year}{2022}), \eprint{2204.13690}.

\bibitem[{\citenamefont{Zhou et~al.}(2022)\citenamefont{Zhou, Han, and
  Wang}}]{Zhou:2022cql}
\bibinfo{author}{\bibfnamefont{Q.}~\bibnamefont{Zhou}},
  \bibinfo{author}{\bibfnamefont{X.-F.} \bibnamefont{Han}}, \bibnamefont{and}
  \bibinfo{author}{\bibfnamefont{L.}~\bibnamefont{Wang}},
  \bibinfo{journal}{Eur. Phys. J. C} \textbf{\bibinfo{volume}{82}},
  \bibinfo{pages}{1135} (\bibinfo{year}{2022}), \eprint{2204.13027}.

\bibitem[{\citenamefont{Cai et~al.}(2022)\citenamefont{Cai, Qiu, Tang, Yu, and
  Zhang}}]{Cai:2022cti}
\bibinfo{author}{\bibfnamefont{C.}~\bibnamefont{Cai}},
  \bibinfo{author}{\bibfnamefont{D.}~\bibnamefont{Qiu}},
  \bibinfo{author}{\bibfnamefont{Y.-L.} \bibnamefont{Tang}},
  \bibinfo{author}{\bibfnamefont{Z.-H.} \bibnamefont{Yu}}, \bibnamefont{and}
  \bibinfo{author}{\bibfnamefont{H.-H.} \bibnamefont{Zhang}},
  \bibinfo{journal}{Phys. Rev. D} \textbf{\bibinfo{volume}{106}},
  \bibinfo{pages}{095003} (\bibinfo{year}{2022}), \eprint{2204.11570}.

\bibitem[{\citenamefont{Batra et~al.}(2022{\natexlab{a}})\citenamefont{Batra,
  A, Mandal, Prajapati, and Srivastava}}]{Batra:2022pej}
\bibinfo{author}{\bibfnamefont{A.}~\bibnamefont{Batra}},
  \bibinfo{author}{\bibfnamefont{S.~K.} \bibnamefont{A}},
  \bibinfo{author}{\bibfnamefont{S.}~\bibnamefont{Mandal}},
  \bibinfo{author}{\bibfnamefont{H.}~\bibnamefont{Prajapati}},
  \bibnamefont{and}
  \bibinfo{author}{\bibfnamefont{R.}~\bibnamefont{Srivastava}}
  (\bibinfo{year}{2022}{\natexlab{a}}), \eprint{2204.11945}.

\bibitem[{\citenamefont{Abouabid et~al.}(2022)\citenamefont{Abouabid, Arhrib,
  Benbrik, Krab, and Ouchemhou}}]{Abouabid:2022lpg}
\bibinfo{author}{\bibfnamefont{H.}~\bibnamefont{Abouabid}},
  \bibinfo{author}{\bibfnamefont{A.}~\bibnamefont{Arhrib}},
  \bibinfo{author}{\bibfnamefont{R.}~\bibnamefont{Benbrik}},
  \bibinfo{author}{\bibfnamefont{M.}~\bibnamefont{Krab}}, \bibnamefont{and}
  \bibinfo{author}{\bibfnamefont{M.}~\bibnamefont{Ouchemhou}}
  (\bibinfo{year}{2022}), \eprint{2204.12018}.

\bibitem[{\citenamefont{Heeck}(2022)}]{Heeck:2022fvl}
\bibinfo{author}{\bibfnamefont{J.}~\bibnamefont{Heeck}},
  \bibinfo{journal}{Phys. Rev. D} \textbf{\bibinfo{volume}{106}},
  \bibinfo{pages}{015004} (\bibinfo{year}{2022}), \eprint{2204.10274}.

\bibitem[{\citenamefont{Addazi et~al.}(2022)\citenamefont{Addazi, Marciano,
  Morais, Pasechnik, and Yang}}]{Addazi:2022fbj}
\bibinfo{author}{\bibfnamefont{A.}~\bibnamefont{Addazi}},
  \bibinfo{author}{\bibfnamefont{A.}~\bibnamefont{Marciano}},
  \bibinfo{author}{\bibfnamefont{A.~P.} \bibnamefont{Morais}},
  \bibinfo{author}{\bibfnamefont{R.}~\bibnamefont{Pasechnik}},
  \bibnamefont{and} \bibinfo{author}{\bibfnamefont{H.}~\bibnamefont{Yang}}
  (\bibinfo{year}{2022}), \eprint{2204.10315}.

\bibitem[{\citenamefont{Cheng et~al.}(2022)\citenamefont{Cheng, He, Huang, Sun,
  and Xing}}]{Cheng:2022aau}
\bibinfo{author}{\bibfnamefont{Y.}~\bibnamefont{Cheng}},
  \bibinfo{author}{\bibfnamefont{X.-G.} \bibnamefont{He}},
  \bibinfo{author}{\bibfnamefont{F.}~\bibnamefont{Huang}},
  \bibinfo{author}{\bibfnamefont{J.}~\bibnamefont{Sun}}, \bibnamefont{and}
  \bibinfo{author}{\bibfnamefont{Z.-P.} \bibnamefont{Xing}},
  \bibinfo{journal}{Phys. Rev. D} \textbf{\bibinfo{volume}{106}},
  \bibinfo{pages}{055011} (\bibinfo{year}{2022}), \eprint{2204.10156}.

\bibitem[{\citenamefont{Batra et~al.}(2022{\natexlab{b}})\citenamefont{Batra,
  K.~A., Mandal, and Srivastava}}]{Batra:2022org}
\bibinfo{author}{\bibfnamefont{A.}~\bibnamefont{Batra}},
  \bibinfo{author}{\bibfnamefont{S.}~\bibnamefont{K.~A.}},
  \bibinfo{author}{\bibfnamefont{S.}~\bibnamefont{Mandal}}, \bibnamefont{and}
  \bibinfo{author}{\bibfnamefont{R.}~\bibnamefont{Srivastava}}
  (\bibinfo{year}{2022}{\natexlab{b}}), \eprint{2204.09376}.

\bibitem[{\citenamefont{Borah et~al.}(2022)\citenamefont{Borah, Mahapatra, and
  Sahu}}]{Borah:2022zim}
\bibinfo{author}{\bibfnamefont{D.}~\bibnamefont{Borah}},
  \bibinfo{author}{\bibfnamefont{S.}~\bibnamefont{Mahapatra}},
  \bibnamefont{and} \bibinfo{author}{\bibfnamefont{N.}~\bibnamefont{Sahu}},
  \bibinfo{journal}{Phys. Lett. B} \textbf{\bibinfo{volume}{831}},
  \bibinfo{pages}{137196} (\bibinfo{year}{2022}), \eprint{2204.09671}.

\bibitem[{\citenamefont{Cao et~al.}(2022)\citenamefont{Cao, Meng, Shang, Wang,
  and Yang}}]{Cao:2022mif}
\bibinfo{author}{\bibfnamefont{J.}~\bibnamefont{Cao}},
  \bibinfo{author}{\bibfnamefont{L.}~\bibnamefont{Meng}},
  \bibinfo{author}{\bibfnamefont{L.}~\bibnamefont{Shang}},
  \bibinfo{author}{\bibfnamefont{S.}~\bibnamefont{Wang}}, \bibnamefont{and}
  \bibinfo{author}{\bibfnamefont{B.}~\bibnamefont{Yang}},
  \bibinfo{journal}{Phys. Rev. D} \textbf{\bibinfo{volume}{106}},
  \bibinfo{pages}{055042} (\bibinfo{year}{2022}), \eprint{2204.09477}.

\bibitem[{\citenamefont{Bhaskar et~al.}(2022)\citenamefont{Bhaskar, Madathil,
  Mandal, and Mitra}}]{Bhaskar:2022vgk}
\bibinfo{author}{\bibfnamefont{A.}~\bibnamefont{Bhaskar}},
  \bibinfo{author}{\bibfnamefont{A.~A.} \bibnamefont{Madathil}},
  \bibinfo{author}{\bibfnamefont{T.}~\bibnamefont{Mandal}}, \bibnamefont{and}
  \bibinfo{author}{\bibfnamefont{S.}~\bibnamefont{Mitra}},
  \bibinfo{journal}{Phys. Rev. D} \textbf{\bibinfo{volume}{106}},
  \bibinfo{pages}{115009} (\bibinfo{year}{2022}), \eprint{2204.09031}.

\bibitem[{\citenamefont{Baek}(2022)}]{Baek:2022agi}
\bibinfo{author}{\bibfnamefont{S.}~\bibnamefont{Baek}} (\bibinfo{year}{2022}),
  \eprint{2204.09585}.

\bibitem[{\citenamefont{Chowdhury et~al.}(2022)\citenamefont{Chowdhury, Heeck,
  Thapa, and Saad}}]{Chowdhury:2022moc}
\bibinfo{author}{\bibfnamefont{T.~A.} \bibnamefont{Chowdhury}},
  \bibinfo{author}{\bibfnamefont{J.}~\bibnamefont{Heeck}},
  \bibinfo{author}{\bibfnamefont{A.}~\bibnamefont{Thapa}}, \bibnamefont{and}
  \bibinfo{author}{\bibfnamefont{S.}~\bibnamefont{Saad}},
  \bibinfo{journal}{Phys. Rev. D} \textbf{\bibinfo{volume}{106}},
  \bibinfo{pages}{035004} (\bibinfo{year}{2022}), \eprint{2204.08390}.

\bibitem[{\citenamefont{Kanemura and Yagyu}(2022)}]{Kanemura:2022ahw}
\bibinfo{author}{\bibfnamefont{S.}~\bibnamefont{Kanemura}} \bibnamefont{and}
  \bibinfo{author}{\bibfnamefont{K.}~\bibnamefont{Yagyu}},
  \bibinfo{journal}{Phys. Lett. B} \textbf{\bibinfo{volume}{831}},
  \bibinfo{pages}{137217} (\bibinfo{year}{2022}), \eprint{2204.07511}.

\bibitem[{\citenamefont{Popov and Srivastava}(2022)}]{Popov:2022ldh}
\bibinfo{author}{\bibfnamefont{O.}~\bibnamefont{Popov}} \bibnamefont{and}
  \bibinfo{author}{\bibfnamefont{R.}~\bibnamefont{Srivastava}}
  (\bibinfo{year}{2022}), \eprint{2204.08568}.

\bibitem[{\citenamefont{Kawamura et~al.}(2022)\citenamefont{Kawamura, Okawa,
  and Omura}}]{Kawamura:2022uft}
\bibinfo{author}{\bibfnamefont{J.}~\bibnamefont{Kawamura}},
  \bibinfo{author}{\bibfnamefont{S.}~\bibnamefont{Okawa}}, \bibnamefont{and}
  \bibinfo{author}{\bibfnamefont{Y.}~\bibnamefont{Omura}},
  \bibinfo{journal}{Phys. Rev. D} \textbf{\bibinfo{volume}{106}},
  \bibinfo{pages}{015005} (\bibinfo{year}{2022}), \eprint{2204.07022}.

\bibitem[{\citenamefont{Ghoshal et~al.}(2022)\citenamefont{Ghoshal, Okada,
  Okada, Raut, Shafi, and Thapa}}]{Ghoshal:2022vzo}
\bibinfo{author}{\bibfnamefont{A.}~\bibnamefont{Ghoshal}},
  \bibinfo{author}{\bibfnamefont{N.}~\bibnamefont{Okada}},
  \bibinfo{author}{\bibfnamefont{S.}~\bibnamefont{Okada}},
  \bibinfo{author}{\bibfnamefont{D.}~\bibnamefont{Raut}},
  \bibinfo{author}{\bibfnamefont{Q.}~\bibnamefont{Shafi}}, \bibnamefont{and}
  \bibinfo{author}{\bibfnamefont{A.}~\bibnamefont{Thapa}}
  (\bibinfo{year}{2022}), \eprint{2204.07138}.

\bibitem[{\citenamefont{Ahn et~al.}(2022)\citenamefont{Ahn, Kang, and
  Ramos}}]{Ahn:2022xax}
\bibinfo{author}{\bibfnamefont{Y.~H.} \bibnamefont{Ahn}},
  \bibinfo{author}{\bibfnamefont{S.~K.} \bibnamefont{Kang}}, \bibnamefont{and}
  \bibinfo{author}{\bibfnamefont{R.}~\bibnamefont{Ramos}},
  \bibinfo{journal}{Phys. Rev. D} \textbf{\bibinfo{volume}{106}},
  \bibinfo{pages}{055038} (\bibinfo{year}{2022}), \eprint{2204.06485}.

\bibitem[{\citenamefont{Balkin et~al.}(2022)\citenamefont{Balkin, Madge, Menzo,
  Perez, Soreq, and Zupan}}]{Balkin:2022glu}
\bibinfo{author}{\bibfnamefont{R.}~\bibnamefont{Balkin}},
  \bibinfo{author}{\bibfnamefont{E.}~\bibnamefont{Madge}},
  \bibinfo{author}{\bibfnamefont{T.}~\bibnamefont{Menzo}},
  \bibinfo{author}{\bibfnamefont{G.}~\bibnamefont{Perez}},
  \bibinfo{author}{\bibfnamefont{Y.}~\bibnamefont{Soreq}}, \bibnamefont{and}
  \bibinfo{author}{\bibfnamefont{J.}~\bibnamefont{Zupan}},
  \bibinfo{journal}{JHEP} \textbf{\bibinfo{volume}{05}}, \bibinfo{pages}{133}
  (\bibinfo{year}{2022}), \eprint{2204.05992}.

\bibitem[{\citenamefont{Biek\"otter et~al.}(2022)\citenamefont{Biek\"otter,
  Heinemeyer, and Weiglein}}]{Biekotter:2022abc}
\bibinfo{author}{\bibfnamefont{T.}~\bibnamefont{Biek\"otter}},
  \bibinfo{author}{\bibfnamefont{S.}~\bibnamefont{Heinemeyer}},
  \bibnamefont{and} \bibinfo{author}{\bibfnamefont{G.}~\bibnamefont{Weiglein}}
  (\bibinfo{year}{2022}), \eprint{2204.05975}.

\bibitem[{\citenamefont{Paul and Valli}(2022)}]{Paul:2022dds}
\bibinfo{author}{\bibfnamefont{A.}~\bibnamefont{Paul}} \bibnamefont{and}
  \bibinfo{author}{\bibfnamefont{M.}~\bibnamefont{Valli}},
  \bibinfo{journal}{Phys. Rev. D} \textbf{\bibinfo{volume}{106}},
  \bibinfo{pages}{013008} (\bibinfo{year}{2022}), \eprint{2204.05267}.

\bibitem[{\citenamefont{Babu et~al.}(2022)\citenamefont{Babu, Jana, and
  K.}}]{Babu:2022pdn}
\bibinfo{author}{\bibfnamefont{K.~S.} \bibnamefont{Babu}},
  \bibinfo{author}{\bibfnamefont{S.}~\bibnamefont{Jana}}, \bibnamefont{and}
  \bibinfo{author}{\bibfnamefont{V.~P.} \bibnamefont{K.}},
  \bibinfo{journal}{Phys. Rev. Lett.} \textbf{\bibinfo{volume}{129}},
  \bibinfo{pages}{121803} (\bibinfo{year}{2022}), \eprint{2204.05303}.

\bibitem[{\citenamefont{Gu et~al.}(2022)\citenamefont{Gu, Liu, Ma, and
  Shu}}]{Gu:2022htv}
\bibinfo{author}{\bibfnamefont{J.}~\bibnamefont{Gu}},
  \bibinfo{author}{\bibfnamefont{Z.}~\bibnamefont{Liu}},
  \bibinfo{author}{\bibfnamefont{T.}~\bibnamefont{Ma}}, \bibnamefont{and}
  \bibinfo{author}{\bibfnamefont{J.}~\bibnamefont{Shu}},
  \bibinfo{journal}{Chin. Phys. C} \textbf{\bibinfo{volume}{46}},
  \bibinfo{pages}{123107} (\bibinfo{year}{2022}), \eprint{2204.05296}.

\bibitem[{\citenamefont{Di~Luzio et~al.}(2022)\citenamefont{Di~Luzio, Gr\"ober,
  and Paradisi}}]{DiLuzio:2022xns}
\bibinfo{author}{\bibfnamefont{L.}~\bibnamefont{Di~Luzio}},
  \bibinfo{author}{\bibfnamefont{R.}~\bibnamefont{Gr\"ober}}, \bibnamefont{and}
  \bibinfo{author}{\bibfnamefont{P.}~\bibnamefont{Paradisi}},
  \bibinfo{journal}{Phys. Lett. B} \textbf{\bibinfo{volume}{832}},
  \bibinfo{pages}{137250} (\bibinfo{year}{2022}), \eprint{2204.05284}.

\bibitem[{\citenamefont{Sakurai et~al.}(2022)\citenamefont{Sakurai, Takahashi,
  and Yin}}]{Sakurai:2022hwh}
\bibinfo{author}{\bibfnamefont{K.}~\bibnamefont{Sakurai}},
  \bibinfo{author}{\bibfnamefont{F.}~\bibnamefont{Takahashi}},
  \bibnamefont{and} \bibinfo{author}{\bibfnamefont{W.}~\bibnamefont{Yin}},
  \bibinfo{journal}{Phys. Lett. B} \textbf{\bibinfo{volume}{833}},
  \bibinfo{pages}{137324} (\bibinfo{year}{2022}), \eprint{2204.04770}.

\bibitem[{\citenamefont{Athron et~al.}(2022{\natexlab{a}})\citenamefont{Athron,
  Fowlie, Lu, Wu, Wu, and Zhu}}]{Athron:2022qpo}
\bibinfo{author}{\bibfnamefont{P.}~\bibnamefont{Athron}},
  \bibinfo{author}{\bibfnamefont{A.}~\bibnamefont{Fowlie}},
  \bibinfo{author}{\bibfnamefont{C.-T.} \bibnamefont{Lu}},
  \bibinfo{author}{\bibfnamefont{L.}~\bibnamefont{Wu}},
  \bibinfo{author}{\bibfnamefont{Y.}~\bibnamefont{Wu}}, \bibnamefont{and}
  \bibinfo{author}{\bibfnamefont{B.}~\bibnamefont{Zhu}}
  (\bibinfo{year}{2022}{\natexlab{a}}), \eprint{2204.03996}.

\bibitem[{\citenamefont{Athron et~al.}(2022{\natexlab{b}})\citenamefont{Athron,
  Bach, Jacob, Kotlarski, St\"ockinger, and Voigt}}]{Athron:2022isz}
\bibinfo{author}{\bibfnamefont{P.}~\bibnamefont{Athron}},
  \bibinfo{author}{\bibfnamefont{M.}~\bibnamefont{Bach}},
  \bibinfo{author}{\bibfnamefont{D.~H.~J.} \bibnamefont{Jacob}},
  \bibinfo{author}{\bibfnamefont{W.}~\bibnamefont{Kotlarski}},
  \bibinfo{author}{\bibfnamefont{D.}~\bibnamefont{St\"ockinger}},
  \bibnamefont{and} \bibinfo{author}{\bibfnamefont{A.}~\bibnamefont{Voigt}},
  \bibinfo{journal}{Phys. Rev. D} \textbf{\bibinfo{volume}{106}},
  \bibinfo{pages}{095023} (\bibinfo{year}{2022}{\natexlab{b}}),
  \eprint{2204.05285}.

\bibitem[{\citenamefont{Han et~al.}(2022)\citenamefont{Han, Wang, Wang, Yang,
  and Zhang}}]{Han:2022juu}
\bibinfo{author}{\bibfnamefont{X.-F.} \bibnamefont{Han}},
  \bibinfo{author}{\bibfnamefont{F.}~\bibnamefont{Wang}},
  \bibinfo{author}{\bibfnamefont{L.}~\bibnamefont{Wang}},
  \bibinfo{author}{\bibfnamefont{J.~M.} \bibnamefont{Yang}}, \bibnamefont{and}
  \bibinfo{author}{\bibfnamefont{Y.}~\bibnamefont{Zhang}},
  \bibinfo{journal}{Chin. Phys. C} \textbf{\bibinfo{volume}{46}},
  \bibinfo{pages}{103105} (\bibinfo{year}{2022}), \eprint{2204.06505}.

\bibitem[{\citenamefont{Basiouris and Leontaris}(2022)}]{Basiouris:2022wei}
\bibinfo{author}{\bibfnamefont{V.}~\bibnamefont{Basiouris}} \bibnamefont{and}
  \bibinfo{author}{\bibfnamefont{G.~K.} \bibnamefont{Leontaris}},
  \bibinfo{journal}{Eur. Phys. J. C} \textbf{\bibinfo{volume}{82}},
  \bibinfo{pages}{1041} (\bibinfo{year}{2022}), \eprint{2205.00758}.

\bibitem[{\citenamefont{Peskin and Takeuchi}(1992)}]{Peskin:1991sw}
\bibinfo{author}{\bibfnamefont{M.~E.} \bibnamefont{Peskin}} \bibnamefont{and}
  \bibinfo{author}{\bibfnamefont{T.}~\bibnamefont{Takeuchi}},
  \bibinfo{journal}{Phys. Rev. D} \textbf{\bibinfo{volume}{46}},
  \bibinfo{pages}{381} (\bibinfo{year}{1992}).

\bibitem[{\citenamefont{Dor\v{s}ner et~al.}(2016)\citenamefont{Dor\v{s}ner,
  Fajfer, Greljo, Kamenik, and Ko\v{s}nik}}]{Dorsner:2016wpm}
\bibinfo{author}{\bibfnamefont{I.}~\bibnamefont{Dor\v{s}ner}},
  \bibinfo{author}{\bibfnamefont{S.}~\bibnamefont{Fajfer}},
  \bibinfo{author}{\bibfnamefont{A.}~\bibnamefont{Greljo}},
  \bibinfo{author}{\bibfnamefont{J.~F.} \bibnamefont{Kamenik}},
  \bibnamefont{and}
  \bibinfo{author}{\bibfnamefont{N.}~\bibnamefont{Ko\v{s}nik}},
  \bibinfo{journal}{Phys. Rept.} \textbf{\bibinfo{volume}{641}},
  \bibinfo{pages}{1} (\bibinfo{year}{2016}), \eprint{1603.04993}.

\bibitem[{\citenamefont{Angelescu et~al.}(2018)\citenamefont{Angelescu,
  Be\v{c}irevi\'c, Faroughy, and Sumensari}}]{Angelescu:2018tyl}
\bibinfo{author}{\bibfnamefont{A.}~\bibnamefont{Angelescu}},
  \bibinfo{author}{\bibfnamefont{D.}~\bibnamefont{Be\v{c}irevi\'c}},
  \bibinfo{author}{\bibfnamefont{D.~A.} \bibnamefont{Faroughy}},
  \bibnamefont{and}
  \bibinfo{author}{\bibfnamefont{O.}~\bibnamefont{Sumensari}},
  \bibinfo{journal}{JHEP} \textbf{\bibinfo{volume}{10}}, \bibinfo{pages}{183}
  (\bibinfo{year}{2018}), \eprint{1808.08179}.

\bibitem[{\citenamefont{Crivellin et~al.}(2022)\citenamefont{Crivellin, Fuks,
  and Schnell}}]{Crivellin:2022mff}
\bibinfo{author}{\bibfnamefont{A.}~\bibnamefont{Crivellin}},
  \bibinfo{author}{\bibfnamefont{B.}~\bibnamefont{Fuks}}, \bibnamefont{and}
  \bibinfo{author}{\bibfnamefont{L.}~\bibnamefont{Schnell}},
  \bibinfo{journal}{JHEP} \textbf{\bibinfo{volume}{06}}, \bibinfo{pages}{169}
  (\bibinfo{year}{2022}), \eprint{2203.10111}.

\bibitem[{\citenamefont{Crivellin
  et~al.}(2020{\natexlab{a}})\citenamefont{Crivellin, M\"uller, and
  Saturnino}}]{Crivellin:2019dwb}
\bibinfo{author}{\bibfnamefont{A.}~\bibnamefont{Crivellin}},
  \bibinfo{author}{\bibfnamefont{D.}~\bibnamefont{M\"uller}}, \bibnamefont{and}
  \bibinfo{author}{\bibfnamefont{F.}~\bibnamefont{Saturnino}},
  \bibinfo{journal}{JHEP} \textbf{\bibinfo{volume}{06}}, \bibinfo{pages}{020}
  (\bibinfo{year}{2020}{\natexlab{a}}), \eprint{1912.04224}.

\bibitem[{\citenamefont{Crivellin and Saturnino}(2019)}]{Crivellin:2019szf}
\bibinfo{author}{\bibfnamefont{A.}~\bibnamefont{Crivellin}} \bibnamefont{and}
  \bibinfo{author}{\bibfnamefont{F.}~\bibnamefont{Saturnino}},
  \bibinfo{journal}{PoS} \textbf{\bibinfo{volume}{DIS2019}},
  \bibinfo{pages}{163} (\bibinfo{year}{2019}), \eprint{1906.01222}.

\bibitem[{\citenamefont{Allanach and Davighi}(2022)}]{Allanach:2022iod}
\bibinfo{author}{\bibfnamefont{B.}~\bibnamefont{Allanach}} \bibnamefont{and}
  \bibinfo{author}{\bibfnamefont{J.}~\bibnamefont{Davighi}}
  (\bibinfo{year}{2022}), \eprint{2211.11766}.

\bibitem[{\citenamefont{Sahoo et~al.}(2021)\citenamefont{Sahoo, Singirala, and
  Mohanta}}]{Sahoo:2021vug}
\bibinfo{author}{\bibfnamefont{S.}~\bibnamefont{Sahoo}},
  \bibinfo{author}{\bibfnamefont{S.}~\bibnamefont{Singirala}},
  \bibnamefont{and} \bibinfo{author}{\bibfnamefont{R.}~\bibnamefont{Mohanta}}
  (\bibinfo{year}{2021}), \eprint{2112.04382}.

\bibitem[{\citenamefont{Belanger et~al.}(2022)}]{Belanger:2021smw}
\bibinfo{author}{\bibfnamefont{G.}~\bibnamefont{Belanger}}
  \bibnamefont{et~al.}, \bibinfo{journal}{JHEP} \textbf{\bibinfo{volume}{02}},
  \bibinfo{pages}{042} (\bibinfo{year}{2022}), \eprint{2111.08027}.

\bibitem[{\citenamefont{Nomura and Okada}(2021)}]{Nomura:2021oeu}
\bibinfo{author}{\bibfnamefont{T.}~\bibnamefont{Nomura}} \bibnamefont{and}
  \bibinfo{author}{\bibfnamefont{H.}~\bibnamefont{Okada}},
  \bibinfo{journal}{Phys. Rev. D} \textbf{\bibinfo{volume}{104}},
  \bibinfo{pages}{035042} (\bibinfo{year}{2021}), \eprint{2104.03248}.

\bibitem[{\citenamefont{Angelescu et~al.}(2021)\citenamefont{Angelescu,
  Be\v{c}irevi\'c, Faroughy, Jaffredo, and Sumensari}}]{Angelescu:2021lln}
\bibinfo{author}{\bibfnamefont{A.}~\bibnamefont{Angelescu}},
  \bibinfo{author}{\bibfnamefont{D.}~\bibnamefont{Be\v{c}irevi\'c}},
  \bibinfo{author}{\bibfnamefont{D.~A.} \bibnamefont{Faroughy}},
  \bibinfo{author}{\bibfnamefont{F.}~\bibnamefont{Jaffredo}}, \bibnamefont{and}
  \bibinfo{author}{\bibfnamefont{O.}~\bibnamefont{Sumensari}},
  \bibinfo{journal}{Phys. Rev. D} \textbf{\bibinfo{volume}{104}},
  \bibinfo{pages}{055017} (\bibinfo{year}{2021}), \eprint{2103.12504}.

\bibitem[{\citenamefont{Altmannshofer et~al.}(2020)\citenamefont{Altmannshofer,
  Gori, Patel, Profumo, and Tuckler}}]{Altmannshofer:2020ywf}
\bibinfo{author}{\bibfnamefont{W.}~\bibnamefont{Altmannshofer}},
  \bibinfo{author}{\bibfnamefont{S.}~\bibnamefont{Gori}},
  \bibinfo{author}{\bibfnamefont{H.~H.} \bibnamefont{Patel}},
  \bibinfo{author}{\bibfnamefont{S.}~\bibnamefont{Profumo}}, \bibnamefont{and}
  \bibinfo{author}{\bibfnamefont{D.}~\bibnamefont{Tuckler}},
  \bibinfo{journal}{JHEP} \textbf{\bibinfo{volume}{05}}, \bibinfo{pages}{069}
  (\bibinfo{year}{2020}), \eprint{2002.01400}.

\bibitem[{\citenamefont{Aydemir et~al.}(2020)\citenamefont{Aydemir, Mandal, and
  Mitra}}]{Aydemir:2019ynb}
\bibinfo{author}{\bibfnamefont{U.}~\bibnamefont{Aydemir}},
  \bibinfo{author}{\bibfnamefont{T.}~\bibnamefont{Mandal}}, \bibnamefont{and}
  \bibinfo{author}{\bibfnamefont{S.}~\bibnamefont{Mitra}},
  \bibinfo{journal}{Phys. Rev. D} \textbf{\bibinfo{volume}{101}},
  \bibinfo{pages}{015011} (\bibinfo{year}{2020}), \eprint{1902.08108}.

\bibitem[{\citenamefont{Di~Luzio et~al.}(2017)\citenamefont{Di~Luzio, Greljo,
  and Nardecchia}}]{DiLuzio:2017vat}
\bibinfo{author}{\bibfnamefont{L.}~\bibnamefont{Di~Luzio}},
  \bibinfo{author}{\bibfnamefont{A.}~\bibnamefont{Greljo}}, \bibnamefont{and}
  \bibinfo{author}{\bibfnamefont{M.}~\bibnamefont{Nardecchia}},
  \bibinfo{journal}{Phys. Rev. D} \textbf{\bibinfo{volume}{96}},
  \bibinfo{pages}{115011} (\bibinfo{year}{2017}), \eprint{1708.08450}.

\bibitem[{\citenamefont{Cai et~al.}(2017)\citenamefont{Cai, Gargalionis,
  Schmidt, and Volkas}}]{Cai:2017wry}
\bibinfo{author}{\bibfnamefont{Y.}~\bibnamefont{Cai}},
  \bibinfo{author}{\bibfnamefont{J.}~\bibnamefont{Gargalionis}},
  \bibinfo{author}{\bibfnamefont{M.~A.} \bibnamefont{Schmidt}},
  \bibnamefont{and} \bibinfo{author}{\bibfnamefont{R.~R.}
  \bibnamefont{Volkas}}, \bibinfo{journal}{JHEP} \textbf{\bibinfo{volume}{10}},
  \bibinfo{pages}{047} (\bibinfo{year}{2017}), \eprint{1704.05849}.

\bibitem[{\citenamefont{Popov and White}(2017)}]{Popov:2016fzr}
\bibinfo{author}{\bibfnamefont{O.}~\bibnamefont{Popov}} \bibnamefont{and}
  \bibinfo{author}{\bibfnamefont{G.~A.} \bibnamefont{White}},
  \bibinfo{journal}{Nucl. Phys. B} \textbf{\bibinfo{volume}{923}},
  \bibinfo{pages}{324} (\bibinfo{year}{2017}), \eprint{1611.04566}.

\bibitem[{\citenamefont{Li et~al.}(2016)\citenamefont{Li, Yang, and
  Zhang}}]{Li:2016vvp}
\bibinfo{author}{\bibfnamefont{X.-Q.} \bibnamefont{Li}},
  \bibinfo{author}{\bibfnamefont{Y.-D.} \bibnamefont{Yang}}, \bibnamefont{and}
  \bibinfo{author}{\bibfnamefont{X.}~\bibnamefont{Zhang}},
  \bibinfo{journal}{JHEP} \textbf{\bibinfo{volume}{08}}, \bibinfo{pages}{054}
  (\bibinfo{year}{2016}), \eprint{1605.09308}.

\bibitem[{\citenamefont{Be\v{c}irevi\'c
  et~al.}(2016)\citenamefont{Be\v{c}irevi\'c, Fajfer, Ko\v{s}nik, and
  Sumensari}}]{Becirevic:2016yqi}
\bibinfo{author}{\bibfnamefont{D.}~\bibnamefont{Be\v{c}irevi\'c}},
  \bibinfo{author}{\bibfnamefont{S.}~\bibnamefont{Fajfer}},
  \bibinfo{author}{\bibfnamefont{N.}~\bibnamefont{Ko\v{s}nik}},
  \bibnamefont{and}
  \bibinfo{author}{\bibfnamefont{O.}~\bibnamefont{Sumensari}},
  \bibinfo{journal}{Phys. Rev. D} \textbf{\bibinfo{volume}{94}},
  \bibinfo{pages}{115021} (\bibinfo{year}{2016}), \eprint{1608.08501}.

\bibitem[{\citenamefont{Bauer and Neubert}(2016)}]{Bauer:2015knc}
\bibinfo{author}{\bibfnamefont{M.}~\bibnamefont{Bauer}} \bibnamefont{and}
  \bibinfo{author}{\bibfnamefont{M.}~\bibnamefont{Neubert}},
  \bibinfo{journal}{Phys. Rev. Lett.} \textbf{\bibinfo{volume}{116}},
  \bibinfo{pages}{141802} (\bibinfo{year}{2016}), \eprint{1511.01900}.

\bibitem[{\citenamefont{Delle~Rose et~al.}(2020)\citenamefont{Delle~Rose,
  Marzo, and Marzola}}]{DelleRose:2020qak}
\bibinfo{author}{\bibfnamefont{L.}~\bibnamefont{Delle~Rose}},
  \bibinfo{author}{\bibfnamefont{C.}~\bibnamefont{Marzo}}, \bibnamefont{and}
  \bibinfo{author}{\bibfnamefont{L.}~\bibnamefont{Marzola}},
  \bibinfo{journal}{Phys. Rev. D} \textbf{\bibinfo{volume}{102}},
  \bibinfo{pages}{115020} (\bibinfo{year}{2020}), \eprint{2005.12389}.

\bibitem[{\citenamefont{Dedes and Mantzaropoulos}(2021)}]{Dedes:2021abc}
\bibinfo{author}{\bibfnamefont{A.}~\bibnamefont{Dedes}} \bibnamefont{and}
  \bibinfo{author}{\bibfnamefont{K.}~\bibnamefont{Mantzaropoulos}},
  \bibinfo{journal}{JHEP} \textbf{\bibinfo{volume}{11}}, \bibinfo{pages}{166}
  (\bibinfo{year}{2021}), \eprint{2108.10055}.

\bibitem[{\citenamefont{Coluccio~Leskow
  et~al.}(2017)\citenamefont{Coluccio~Leskow, D'Ambrosio, Crivellin, and
  M\"uller}}]{ColuccioLeskow:2016dox}
\bibinfo{author}{\bibfnamefont{E.}~\bibnamefont{Coluccio~Leskow}},
  \bibinfo{author}{\bibfnamefont{G.}~\bibnamefont{D'Ambrosio}},
  \bibinfo{author}{\bibfnamefont{A.}~\bibnamefont{Crivellin}},
  \bibnamefont{and} \bibinfo{author}{\bibfnamefont{D.}~\bibnamefont{M\"uller}},
  \bibinfo{journal}{Phys. Rev. D} \textbf{\bibinfo{volume}{95}},
  \bibinfo{pages}{055018} (\bibinfo{year}{2017}), \eprint{1612.06858}.

\bibitem[{\citenamefont{Crivellin et~al.}(2021)\citenamefont{Crivellin,
  Mueller, and Saturnino}}]{Crivellin:2020tsz}
\bibinfo{author}{\bibfnamefont{A.}~\bibnamefont{Crivellin}},
  \bibinfo{author}{\bibfnamefont{D.}~\bibnamefont{Mueller}}, \bibnamefont{and}
  \bibinfo{author}{\bibfnamefont{F.}~\bibnamefont{Saturnino}},
  \bibinfo{journal}{Phys. Rev. Lett.} \textbf{\bibinfo{volume}{127}},
  \bibinfo{pages}{021801} (\bibinfo{year}{2021}), \eprint{2008.02643}.

\bibitem[{\citenamefont{Crivellin
  et~al.}(2020{\natexlab{b}})\citenamefont{Crivellin, M\"uller, and
  Saturnino}}]{Crivellin:2020ukd}
\bibinfo{author}{\bibfnamefont{A.}~\bibnamefont{Crivellin}},
  \bibinfo{author}{\bibfnamefont{D.}~\bibnamefont{M\"uller}}, \bibnamefont{and}
  \bibinfo{author}{\bibfnamefont{F.}~\bibnamefont{Saturnino}},
  \bibinfo{journal}{JHEP} \textbf{\bibinfo{volume}{11}}, \bibinfo{pages}{094}
  (\bibinfo{year}{2020}{\natexlab{b}}), \eprint{2006.10758}.

\bibitem[{\citenamefont{Denner}(1993)}]{Denner:1991kt}
\bibinfo{author}{\bibfnamefont{A.}~\bibnamefont{Denner}},
  \bibinfo{journal}{Fortsch. Phys.} \textbf{\bibinfo{volume}{41}},
  \bibinfo{pages}{307} (\bibinfo{year}{1993}), \eprint{0709.1075}.

\bibitem[{\citenamefont{Sirlin}(1980)}]{Sirlin:1980nh}
\bibinfo{author}{\bibfnamefont{A.}~\bibnamefont{Sirlin}},
  \bibinfo{journal}{Phys. Rev. D} \textbf{\bibinfo{volume}{22}},
  \bibinfo{pages}{971} (\bibinfo{year}{1980}).

\bibitem[{\citenamefont{Green and Veltman}(1980)}]{Green:1980bd}
\bibinfo{author}{\bibfnamefont{M.}~\bibnamefont{Green}} \bibnamefont{and}
  \bibinfo{author}{\bibfnamefont{M.~J.~G.} \bibnamefont{Veltman}},
  \bibinfo{journal}{Nucl. Phys. B} \textbf{\bibinfo{volume}{169}},
  \bibinfo{pages}{137} (\bibinfo{year}{1980}), \bibinfo{note}{[Erratum:
  Nucl.Phys.B 175, 547 (1980)]}.

\bibitem[{\citenamefont{Bardin and Passarino}(1999)}]{Bardin:1999ak}
\bibinfo{author}{\bibfnamefont{D.~Y.} \bibnamefont{Bardin}} \bibnamefont{and}
  \bibinfo{author}{\bibfnamefont{G.}~\bibnamefont{Passarino}},
  \emph{\bibinfo{title}{{The standard model in the making: Precision study of
  the electroweak interactions}}} (\bibinfo{year}{1999}).

\bibitem[{\citenamefont{Freitas et~al.}(2003)\citenamefont{Freitas, Heinemeyer,
  and Weiglein}}]{Freitas:2002ve}
\bibinfo{author}{\bibfnamefont{A.}~\bibnamefont{Freitas}},
  \bibinfo{author}{\bibfnamefont{S.}~\bibnamefont{Heinemeyer}},
  \bibnamefont{and} \bibinfo{author}{\bibfnamefont{G.}~\bibnamefont{Weiglein}},
  \bibinfo{journal}{Nucl. Phys. B Proc. Suppl.} \textbf{\bibinfo{volume}{116}},
  \bibinfo{pages}{331} (\bibinfo{year}{2003}), \eprint{hep-ph/0212068}.

\bibitem[{\citenamefont{Awramik et~al.}(2004)\citenamefont{Awramik, Czakon,
  Freitas, and Weiglein}}]{Awramik:2003rn}
\bibinfo{author}{\bibfnamefont{M.}~\bibnamefont{Awramik}},
  \bibinfo{author}{\bibfnamefont{M.}~\bibnamefont{Czakon}},
  \bibinfo{author}{\bibfnamefont{A.}~\bibnamefont{Freitas}}, \bibnamefont{and}
  \bibinfo{author}{\bibfnamefont{G.}~\bibnamefont{Weiglein}},
  \bibinfo{journal}{Phys. Rev. D} \textbf{\bibinfo{volume}{69}},
  \bibinfo{pages}{053006} (\bibinfo{year}{2004}), \eprint{hep-ph/0311148}.

\bibitem[{\citenamefont{Awramik and
  Czakon}(2003{\natexlab{a}})}]{Awramik:2003ee}
\bibinfo{author}{\bibfnamefont{M.}~\bibnamefont{Awramik}} \bibnamefont{and}
  \bibinfo{author}{\bibfnamefont{M.}~\bibnamefont{Czakon}},
  \bibinfo{journal}{Phys. Lett. B} \textbf{\bibinfo{volume}{568}},
  \bibinfo{pages}{48} (\bibinfo{year}{2003}{\natexlab{a}}),
  \eprint{hep-ph/0305248}.

\bibitem[{\citenamefont{Awramik and
  Czakon}(2003{\natexlab{b}})}]{Awramik:2002wv}
\bibinfo{author}{\bibfnamefont{M.}~\bibnamefont{Awramik}} \bibnamefont{and}
  \bibinfo{author}{\bibfnamefont{M.}~\bibnamefont{Czakon}},
  \bibinfo{journal}{Nucl. Phys. B Proc. Suppl.} \textbf{\bibinfo{volume}{116}},
  \bibinfo{pages}{238} (\bibinfo{year}{2003}{\natexlab{b}}),
  \eprint{hep-ph/0211041}.

\bibitem[{\citenamefont{Awramik et~al.}(2003)\citenamefont{Awramik, Czakon,
  Onishchenko, and Veretin}}]{Awramik:2002vu}
\bibinfo{author}{\bibfnamefont{M.}~\bibnamefont{Awramik}},
  \bibinfo{author}{\bibfnamefont{M.}~\bibnamefont{Czakon}},
  \bibinfo{author}{\bibfnamefont{A.}~\bibnamefont{Onishchenko}},
  \bibnamefont{and} \bibinfo{author}{\bibfnamefont{O.}~\bibnamefont{Veretin}},
  \bibinfo{journal}{Phys. Rev. D} \textbf{\bibinfo{volume}{68}},
  \bibinfo{pages}{053004} (\bibinfo{year}{2003}), \eprint{hep-ph/0209084}.

\bibitem[{\citenamefont{Awramik and Czakon}(2002)}]{Awramik:2002wn}
\bibinfo{author}{\bibfnamefont{M.}~\bibnamefont{Awramik}} \bibnamefont{and}
  \bibinfo{author}{\bibfnamefont{M.}~\bibnamefont{Czakon}},
  \bibinfo{journal}{Phys. Rev. Lett.} \textbf{\bibinfo{volume}{89}},
  \bibinfo{pages}{241801} (\bibinfo{year}{2002}), \eprint{hep-ph/0208113}.

\bibitem[{\citenamefont{Freitas et~al.}(2002)\citenamefont{Freitas, Hollik,
  Walter, and Weiglein}}]{Freitas:2002ja}
\bibinfo{author}{\bibfnamefont{A.}~\bibnamefont{Freitas}},
  \bibinfo{author}{\bibfnamefont{W.}~\bibnamefont{Hollik}},
  \bibinfo{author}{\bibfnamefont{W.}~\bibnamefont{Walter}}, \bibnamefont{and}
  \bibinfo{author}{\bibfnamefont{G.}~\bibnamefont{Weiglein}},
  \bibinfo{journal}{Nucl. Phys. B} \textbf{\bibinfo{volume}{632}},
  \bibinfo{pages}{189} (\bibinfo{year}{2002}), \bibinfo{note}{[Erratum:
  Nucl.Phys.B 666, 305--307 (2003)]}, \eprint{hep-ph/0202131}.

\bibitem[{\citenamefont{Freitas et~al.}(2000)\citenamefont{Freitas, Hollik,
  Walter, and Weiglein}}]{Freitas:2000gg}
\bibinfo{author}{\bibfnamefont{A.}~\bibnamefont{Freitas}},
  \bibinfo{author}{\bibfnamefont{W.}~\bibnamefont{Hollik}},
  \bibinfo{author}{\bibfnamefont{W.}~\bibnamefont{Walter}}, \bibnamefont{and}
  \bibinfo{author}{\bibfnamefont{G.}~\bibnamefont{Weiglein}},
  \bibinfo{journal}{Phys. Lett. B} \textbf{\bibinfo{volume}{495}},
  \bibinfo{pages}{338} (\bibinfo{year}{2000}), \bibinfo{note}{[Erratum:
  Phys.Lett.B 570, 265 (2003)]}, \eprint{hep-ph/0007091}.

\bibitem[{\citenamefont{Onishchenko and Veretin}(2003)}]{Onishchenko:2002ve}
\bibinfo{author}{\bibfnamefont{A.}~\bibnamefont{Onishchenko}} \bibnamefont{and}
  \bibinfo{author}{\bibfnamefont{O.}~\bibnamefont{Veretin}},
  \bibinfo{journal}{Phys. Lett. B} \textbf{\bibinfo{volume}{551}},
  \bibinfo{pages}{111} (\bibinfo{year}{2003}), \eprint{hep-ph/0209010}.

\bibitem[{\citenamefont{Avdeev et~al.}(1994)\citenamefont{Avdeev, Fleischer,
  Mikhailov, and Tarasov}}]{Avdeev:1994db}
\bibinfo{author}{\bibfnamefont{L.}~\bibnamefont{Avdeev}},
  \bibinfo{author}{\bibfnamefont{J.}~\bibnamefont{Fleischer}},
  \bibinfo{author}{\bibfnamefont{S.}~\bibnamefont{Mikhailov}},
  \bibnamefont{and} \bibinfo{author}{\bibfnamefont{O.}~\bibnamefont{Tarasov}},
  \bibinfo{journal}{Phys. Lett. B} \textbf{\bibinfo{volume}{336}},
  \bibinfo{pages}{560} (\bibinfo{year}{1994}), \bibinfo{note}{[Erratum:
  Phys.Lett.B 349, 597--598 (1995)]}, \eprint{hep-ph/9406363}.

\bibitem[{\citenamefont{van~der Bij et~al.}(2001)\citenamefont{van~der Bij,
  Chetyrkin, Faisst, Jikia, and Seidensticker}}]{vanderBij:2000cg}
\bibinfo{author}{\bibfnamefont{J.~J.} \bibnamefont{van~der Bij}},
  \bibinfo{author}{\bibfnamefont{K.~G.} \bibnamefont{Chetyrkin}},
  \bibinfo{author}{\bibfnamefont{M.}~\bibnamefont{Faisst}},
  \bibinfo{author}{\bibfnamefont{G.}~\bibnamefont{Jikia}}, \bibnamefont{and}
  \bibinfo{author}{\bibfnamefont{T.}~\bibnamefont{Seidensticker}},
  \bibinfo{journal}{Phys. Lett. B} \textbf{\bibinfo{volume}{498}},
  \bibinfo{pages}{156} (\bibinfo{year}{2001}), \eprint{hep-ph/0011373}.

\bibitem[{\citenamefont{Chetyrkin et~al.}(1996)\citenamefont{Chetyrkin, Kuhn,
  and Steinhauser}}]{Chetyrkin:1996cf}
\bibinfo{author}{\bibfnamefont{K.~G.} \bibnamefont{Chetyrkin}},
  \bibinfo{author}{\bibfnamefont{J.~H.} \bibnamefont{Kuhn}}, \bibnamefont{and}
  \bibinfo{author}{\bibfnamefont{M.}~\bibnamefont{Steinhauser}},
  \bibinfo{journal}{Nucl. Phys. B} \textbf{\bibinfo{volume}{482}},
  \bibinfo{pages}{213} (\bibinfo{year}{1996}), \eprint{hep-ph/9606230}.

\bibitem[{\citenamefont{Chetyrkin
  et~al.}(1995{\natexlab{a}})\citenamefont{Chetyrkin, Kuhn, and
  Steinhauser}}]{Chetyrkin:1995js}
\bibinfo{author}{\bibfnamefont{K.~G.} \bibnamefont{Chetyrkin}},
  \bibinfo{author}{\bibfnamefont{J.~H.} \bibnamefont{Kuhn}}, \bibnamefont{and}
  \bibinfo{author}{\bibfnamefont{M.}~\bibnamefont{Steinhauser}},
  \bibinfo{journal}{Phys. Rev. Lett.} \textbf{\bibinfo{volume}{75}},
  \bibinfo{pages}{3394} (\bibinfo{year}{1995}{\natexlab{a}}),
  \eprint{hep-ph/9504413}.

\bibitem[{\citenamefont{Chetyrkin
  et~al.}(1995{\natexlab{b}})\citenamefont{Chetyrkin, Kuhn, and
  Steinhauser}}]{Chetyrkin:1995ix}
\bibinfo{author}{\bibfnamefont{K.~G.} \bibnamefont{Chetyrkin}},
  \bibinfo{author}{\bibfnamefont{J.~H.} \bibnamefont{Kuhn}}, \bibnamefont{and}
  \bibinfo{author}{\bibfnamefont{M.}~\bibnamefont{Steinhauser}},
  \bibinfo{journal}{Phys. Lett. B} \textbf{\bibinfo{volume}{351}},
  \bibinfo{pages}{331} (\bibinfo{year}{1995}{\natexlab{b}}),
  \eprint{hep-ph/9502291}.

\bibitem[{\citenamefont{Boughezal et~al.}(2005)\citenamefont{Boughezal, Tausk,
  and van~der Bij}}]{Boughezal:2004ef}
\bibinfo{author}{\bibfnamefont{R.}~\bibnamefont{Boughezal}},
  \bibinfo{author}{\bibfnamefont{J.~B.} \bibnamefont{Tausk}}, \bibnamefont{and}
  \bibinfo{author}{\bibfnamefont{J.~J.} \bibnamefont{van~der Bij}},
  \bibinfo{journal}{Nucl. Phys. B} \textbf{\bibinfo{volume}{713}},
  \bibinfo{pages}{278} (\bibinfo{year}{2005}), \eprint{hep-ph/0410216}.

\bibitem[{\citenamefont{Chetyrkin et~al.}(2006)\citenamefont{Chetyrkin, Faisst,
  Kuhn, Maierhofer, and Sturm}}]{Chetyrkin:2006bj}
\bibinfo{author}{\bibfnamefont{K.~G.} \bibnamefont{Chetyrkin}},
  \bibinfo{author}{\bibfnamefont{M.}~\bibnamefont{Faisst}},
  \bibinfo{author}{\bibfnamefont{J.~H.} \bibnamefont{Kuhn}},
  \bibinfo{author}{\bibfnamefont{P.}~\bibnamefont{Maierhofer}},
  \bibnamefont{and} \bibinfo{author}{\bibfnamefont{C.}~\bibnamefont{Sturm}},
  \bibinfo{journal}{Phys. Rev. Lett.} \textbf{\bibinfo{volume}{97}},
  \bibinfo{pages}{102003} (\bibinfo{year}{2006}), \eprint{hep-ph/0605201}.

\bibitem[{\citenamefont{Boughezal and Czakon}(2006)}]{Boughezal:2006xk}
\bibinfo{author}{\bibfnamefont{R.}~\bibnamefont{Boughezal}} \bibnamefont{and}
  \bibinfo{author}{\bibfnamefont{M.}~\bibnamefont{Czakon}},
  \bibinfo{journal}{Nucl. Phys. B} \textbf{\bibinfo{volume}{755}},
  \bibinfo{pages}{221} (\bibinfo{year}{2006}), \eprint{hep-ph/0606232}.

\bibitem[{\citenamefont{Aad et~al.}(2012)}]{ATLAS:2012yve}
\bibinfo{author}{\bibfnamefont{G.}~\bibnamefont{Aad}} \bibnamefont{et~al.}
  (\bibinfo{collaboration}{ATLAS}), \bibinfo{journal}{Phys. Lett. B}
  \textbf{\bibinfo{volume}{716}}, \bibinfo{pages}{1} (\bibinfo{year}{2012}),
  \eprint{1207.7214}.

\bibitem[{\citenamefont{Chatrchyan et~al.}(2012)}]{CMS:2012qbp}
\bibinfo{author}{\bibfnamefont{S.}~\bibnamefont{Chatrchyan}}
  \bibnamefont{et~al.} (\bibinfo{collaboration}{CMS}), \bibinfo{journal}{Phys.
  Lett. B} \textbf{\bibinfo{volume}{716}}, \bibinfo{pages}{30}
  (\bibinfo{year}{2012}), \eprint{1207.7235}.

\bibitem[{\citenamefont{Heinemeyer et~al.}(2013)\citenamefont{Heinemeyer,
  Hollik, Weiglein, and Zeune}}]{Heinemeyer:2013dia}
\bibinfo{author}{\bibfnamefont{S.}~\bibnamefont{Heinemeyer}},
  \bibinfo{author}{\bibfnamefont{W.}~\bibnamefont{Hollik}},
  \bibinfo{author}{\bibfnamefont{G.}~\bibnamefont{Weiglein}}, \bibnamefont{and}
  \bibinfo{author}{\bibfnamefont{L.}~\bibnamefont{Zeune}},
  \bibinfo{journal}{JHEP} \textbf{\bibinfo{volume}{12}}, \bibinfo{pages}{084}
  (\bibinfo{year}{2013}), \eprint{1311.1663}.

\bibitem[{\citenamefont{L\'opez-Val and Robens}(2014)}]{Lopez-Val:2014jva}
\bibinfo{author}{\bibfnamefont{D.}~\bibnamefont{L\'opez-Val}} \bibnamefont{and}
  \bibinfo{author}{\bibfnamefont{T.}~\bibnamefont{Robens}},
  \bibinfo{journal}{Phys. Rev. D} \textbf{\bibinfo{volume}{90}},
  \bibinfo{pages}{114018} (\bibinfo{year}{2014}), \eprint{1406.1043}.

\bibitem[{\citenamefont{Christensen and Duhr}(2009)}]{Christensen:2008py}
\bibinfo{author}{\bibfnamefont{N.~D.} \bibnamefont{Christensen}}
  \bibnamefont{and} \bibinfo{author}{\bibfnamefont{C.}~\bibnamefont{Duhr}},
  \bibinfo{journal}{Comput. Phys. Commun.} \textbf{\bibinfo{volume}{180}},
  \bibinfo{pages}{1614} (\bibinfo{year}{2009}), \eprint{0806.4194}.

\bibitem[{\citenamefont{Alloul et~al.}(2014)\citenamefont{Alloul, Christensen,
  Degrande, Duhr, and Fuks}}]{Alloul:2013bka}
\bibinfo{author}{\bibfnamefont{A.}~\bibnamefont{Alloul}},
  \bibinfo{author}{\bibfnamefont{N.~D.} \bibnamefont{Christensen}},
  \bibinfo{author}{\bibfnamefont{C.}~\bibnamefont{Degrande}},
  \bibinfo{author}{\bibfnamefont{C.}~\bibnamefont{Duhr}}, \bibnamefont{and}
  \bibinfo{author}{\bibfnamefont{B.}~\bibnamefont{Fuks}},
  \bibinfo{journal}{Comput. Phys. Commun.} \textbf{\bibinfo{volume}{185}},
  \bibinfo{pages}{2250} (\bibinfo{year}{2014}), \eprint{1310.1921}.

\bibitem[{\citenamefont{Dor\v{s}ner and Greljo}(2018)}]{Dorsner:2018ynv}
\bibinfo{author}{\bibfnamefont{I.}~\bibnamefont{Dor\v{s}ner}} \bibnamefont{and}
  \bibinfo{author}{\bibfnamefont{A.}~\bibnamefont{Greljo}},
  \bibinfo{journal}{JHEP} \textbf{\bibinfo{volume}{05}}, \bibinfo{pages}{126}
  (\bibinfo{year}{2018}), \eprint{1801.07641}.

\bibitem[{\citenamefont{Hahn}(2000)}]{Hahn:2000jm}
\bibinfo{author}{\bibfnamefont{T.}~\bibnamefont{Hahn}}, \bibinfo{journal}{Nucl.
  Phys. B Proc. Suppl.} \textbf{\bibinfo{volume}{89}}, \bibinfo{pages}{231}
  (\bibinfo{year}{2000}), \eprint{hep-ph/0005029}.

\bibitem[{\citenamefont{Hahn}(2001)}]{Hahn:2000kx}
\bibinfo{author}{\bibfnamefont{T.}~\bibnamefont{Hahn}},
  \bibinfo{journal}{Comput. Phys. Commun.} \textbf{\bibinfo{volume}{140}},
  \bibinfo{pages}{418} (\bibinfo{year}{2001}), \eprint{hep-ph/0012260}.

\bibitem[{\citenamefont{Hahn et~al.}(2016)\citenamefont{Hahn, Pa\ss{}ehr, and
  Schappacher}}]{Hahn:2016ebn}
\bibinfo{author}{\bibfnamefont{T.}~\bibnamefont{Hahn}},
  \bibinfo{author}{\bibfnamefont{S.}~\bibnamefont{Pa\ss{}ehr}},
  \bibnamefont{and}
  \bibinfo{author}{\bibfnamefont{C.}~\bibnamefont{Schappacher}},
  \bibinfo{journal}{PoS} \textbf{\bibinfo{volume}{LL2016}},
  \bibinfo{pages}{068} (\bibinfo{year}{2016}), \eprint{1604.04611}.

\bibitem[{\citenamefont{Patel}(2017)}]{Patel:2016fam}
\bibinfo{author}{\bibfnamefont{H.~H.} \bibnamefont{Patel}},
  \bibinfo{journal}{Comput. Phys. Commun.} \textbf{\bibinfo{volume}{218}},
  \bibinfo{pages}{66} (\bibinfo{year}{2017}), \eprint{1612.00009}.

\bibitem[{\citenamefont{Aaij et~al.}(2022)}]{LHCb:2021bjt}
\bibinfo{author}{\bibfnamefont{R.}~\bibnamefont{Aaij}} \bibnamefont{et~al.}
  (\bibinfo{collaboration}{LHCb}), \bibinfo{journal}{JHEP}
  \textbf{\bibinfo{volume}{01}}, \bibinfo{pages}{036} (\bibinfo{year}{2022}),
  \eprint{2109.01113}.

\bibitem[{\citenamefont{Schael et~al.}(2006)}]{ALEPH:2005ab}
\bibinfo{author}{\bibfnamefont{S.}~\bibnamefont{Schael}} \bibnamefont{et~al.}
  (\bibinfo{collaboration}{ALEPH, DELPHI, L3, OPAL, SLD, LEP Electroweak
  Working Group, SLD Electroweak Group, SLD Heavy Flavour Group}),
  \bibinfo{journal}{Phys. Rept.} \textbf{\bibinfo{volume}{427}},
  \bibinfo{pages}{257} (\bibinfo{year}{2006}), \eprint{hep-ex/0509008}.

\bibitem[{\citenamefont{Abazov et~al.}(2012)}]{D0:2012kms}
\bibinfo{author}{\bibfnamefont{V.~M.} \bibnamefont{Abazov}}
  \bibnamefont{et~al.} (\bibinfo{collaboration}{D0}), \bibinfo{journal}{Phys.
  Rev. Lett.} \textbf{\bibinfo{volume}{108}}, \bibinfo{pages}{151804}
  (\bibinfo{year}{2012}), \eprint{1203.0293}.

\bibitem[{\citenamefont{Aaboud et~al.}(2018)}]{ATLAS:2017rzl}
\bibinfo{author}{\bibfnamefont{M.}~\bibnamefont{Aaboud}} \bibnamefont{et~al.}
  (\bibinfo{collaboration}{ATLAS}), \bibinfo{journal}{Eur. Phys. J. C}
  \textbf{\bibinfo{volume}{78}}, \bibinfo{pages}{110} (\bibinfo{year}{2018}),
  \bibinfo{note}{[Erratum: Eur.Phys.J.C 78, 898 (2018)]}, \eprint{1701.07240}.

\bibitem[{\citenamefont{Bagnaschi et~al.}(2022)\citenamefont{Bagnaschi, Ellis,
  Madigan, Mimasu, Sanz, and You}}]{Bagnaschi:2022whn}
\bibinfo{author}{\bibfnamefont{E.}~\bibnamefont{Bagnaschi}},
  \bibinfo{author}{\bibfnamefont{J.}~\bibnamefont{Ellis}},
  \bibinfo{author}{\bibfnamefont{M.}~\bibnamefont{Madigan}},
  \bibinfo{author}{\bibfnamefont{K.}~\bibnamefont{Mimasu}},
  \bibinfo{author}{\bibfnamefont{V.}~\bibnamefont{Sanz}}, \bibnamefont{and}
  \bibinfo{author}{\bibfnamefont{T.}~\bibnamefont{You}},
  \bibinfo{journal}{JHEP} \textbf{\bibinfo{volume}{08}}, \bibinfo{pages}{308}
  (\bibinfo{year}{2022}), \eprint{2204.05260}.

\bibitem[{\citenamefont{Sirunyan et~al.}(2018{\natexlab{a}})}]{CMS:2018txo}
\bibinfo{author}{\bibfnamefont{A.~M.} \bibnamefont{Sirunyan}}
  \bibnamefont{et~al.} (\bibinfo{collaboration}{CMS}), \bibinfo{journal}{JHEP}
  \textbf{\bibinfo{volume}{07}}, \bibinfo{pages}{115}
  (\bibinfo{year}{2018}{\natexlab{a}}), \eprint{1806.03472}.

\bibitem[{\citenamefont{Aad et~al.}(2020)}]{ATLAS:2020dsk}
\bibinfo{author}{\bibfnamefont{G.}~\bibnamefont{Aad}} \bibnamefont{et~al.}
  (\bibinfo{collaboration}{ATLAS}), \bibinfo{journal}{JHEP}
  \textbf{\bibinfo{volume}{10}}, \bibinfo{pages}{112} (\bibinfo{year}{2020}),
  \eprint{2006.05872}.

\bibitem[{\citenamefont{Sirunyan et~al.}(2018{\natexlab{b}})}]{CMS:2018oaj}
\bibinfo{author}{\bibfnamefont{A.~M.} \bibnamefont{Sirunyan}}
  \bibnamefont{et~al.} (\bibinfo{collaboration}{CMS}), \bibinfo{journal}{Phys.
  Rev. Lett.} \textbf{\bibinfo{volume}{121}}, \bibinfo{pages}{241802}
  (\bibinfo{year}{2018}{\natexlab{b}}), \eprint{1809.05558}.

\bibitem[{\citenamefont{Aaboud et~al.}(2019)}]{ATLAS:2019qpq}
\bibinfo{author}{\bibfnamefont{M.}~\bibnamefont{Aaboud}} \bibnamefont{et~al.}
  (\bibinfo{collaboration}{ATLAS}), \bibinfo{journal}{JHEP}
  \textbf{\bibinfo{volume}{06}}, \bibinfo{pages}{144} (\bibinfo{year}{2019}),
  \eprint{1902.08103}.

\bibitem[{\citenamefont{Aad et~al.}(2021{\natexlab{a}})}]{ATLAS:2020xov}
\bibinfo{author}{\bibfnamefont{G.}~\bibnamefont{Aad}} \bibnamefont{et~al.}
  (\bibinfo{collaboration}{ATLAS}), \bibinfo{journal}{Eur. Phys. J. C}
  \textbf{\bibinfo{volume}{81}}, \bibinfo{pages}{313}
  (\bibinfo{year}{2021}{\natexlab{a}}), \eprint{2010.02098}.

\bibitem[{\citenamefont{Aad et~al.}(2021{\natexlab{b}})}]{ATLAS:2021oiz}
\bibinfo{author}{\bibfnamefont{G.}~\bibnamefont{Aad}} \bibnamefont{et~al.}
  (\bibinfo{collaboration}{ATLAS}), \bibinfo{journal}{JHEP}
  \textbf{\bibinfo{volume}{06}}, \bibinfo{pages}{179}
  (\bibinfo{year}{2021}{\natexlab{b}}), \eprint{2101.11582}.

\bibitem[{\citenamefont{Ellis}(2017)}]{Ellis:2016jkw}
\bibinfo{author}{\bibfnamefont{J.}~\bibnamefont{Ellis}},
  \bibinfo{journal}{Comput. Phys. Commun.} \textbf{\bibinfo{volume}{210}},
  \bibinfo{pages}{103} (\bibinfo{year}{2017}), \eprint{1601.05437}.

\end{thebibliography}

\end{document}